\documentclass[twocolumn,superscriptaddress,floatfix,preprintnumbers,nofootinbib]{revtex4-2}
\usepackage[colorlinks=true,breaklinks=true]{hyperref}
\usepackage[utf8]{inputenc}
\usepackage{float}
\usepackage{subcaption}
\hypersetup{allcolors=[rgb]{0.0 0.0 0.6},linkcolor=[rgb]{0.75 0.05 0.05}}
\usepackage{amsmath,amssymb}
\usepackage{mathtools}
\usepackage[normalem]{ulem}
\usepackage{bm}
\usepackage{physics}
\usepackage{siunitx}
\usepackage{xr} 
\DeclareSIUnit \parsec {pc}
\DeclareSIUnit \years {yr}
\DeclareSIUnit \electronvolt {eV}

\usepackage[dvipsnames]{xcolor}
\usepackage{float}
\usepackage{graphicx}
\usepackage{epsfig}
\usepackage{letltxmacro}
\LetLtxMacro{\oldcite}{\cite}
\renewcommand{\cite}[1]{\mbox{\oldcite{#1}}}

\usepackage{booktabs} 

\usepackage{orcidlink}

\long\def\exclude#1{}

\newcommand{\e}{\mathrm{e}}
\newcommand{\iu}{\mathrm{i}}

\long\def\exclude#1{}

\allowdisplaybreaks

\bibpunct{[}{]}{,}{n}{}{,}

\hypersetup{
    colorlinks=true,
    citecolor=[rgb]{.1, .7, .6},
    linkcolor=[rgb]{.2, .55, .95},
    filecolor=magenta,
    urlcolor=[rgb]{.1, .7, .6},
}

\usepackage{appendix}

\begin{document}

\title{Influence of tides and self-gravity on Ultra-Light Dark Matter Bounds from Dwarf Galaxies}

\author{Andrea Caputo} 
\email{andrea.caputo@cern.ch}
\affiliation{Department of Theoretical Physics, CERN, Esplanade des Particules 1, P.O. Box 1211, Geneva 23, Switzerland}
\affiliation{Dipartimento di Fisica, ``Sapienza'' Universit\`a di Roma \& Sezione INFN Roma1, Piazzale Aldo Moro
5, 00185, Roma, Italy}
\affiliation{Department of Particle Physics and Astrophysics, Weizmann Institute of Science, Rehovot 7610001, Israel}

\author{Luca Teodori}
\email{luca.teodori@iac.es}
\affiliation{Instituto de Astrof\'isica de Canarias, C/ V\'ia L\'actea, s/n E38205, La Laguna, Tenerife, Spain
 } 
\affiliation{Universidad de La Laguna, Departamento de Astrof\'isica, La Laguna, Tenerife,
 	Spain}

\begin{abstract}
Dwarf spheroidal galaxies provide some of the most sensitive astrophysical probes of ultra-light dark matter (ULDM), but the inferred constraints can be affected by two important systematics: tidal interactions with the Milky Way, which reduce ULDM-induced dynamical heating, and stellar self-gravity, which can become relevant if the stellar component was more compact at earlier times. In this work, we attempt to estimate both effects by reconstructing dwarf-galaxy orbital histories in a Milky-Way potential, adopting a simple and approximate tidal-susceptibility diagnostic that we argue provides a conservative description of tidal stripping, and explicitly including stellar self-gravity in our numerical simulations. Within our framework, which we apply to five different dwarf galaxies, we find that ULDM with masses $5\times 10^{-22} \lesssim m/{\rm eV} \lesssim 5\times 10^{-21}$
remains in tension with current data. 
\end{abstract}

\preprint{CERN-TH-2026-065}
\maketitle

\section{Introduction}
Ultra-light Dark Matter (ULDM) is a very appealing and studied dark matter candidate~\cite{Hui2017,Hui:2021tkt,Ferreira:2020fam}, whose defining feature is a low enough particle mass $m$ such that its associated de-Broglie wavelength reaches galactic scales. From a cosmological and astrophysical point of view, ULDM can affect the small scales power spectrum and stellar kinematics and dynamics in galaxies. Many different groups have now explored ULDM simulations ~\cite{Guzman2004,Schive:2014dra,Schive:2014hza,Schwabe:2016rze,Veltmaat:2016rxo,Mocz:2017wlg,Veltmaat:2018dfz,Eggemeier:2019jsu,Chen:2020cef,Schwabe:2020eac,Zhang:2018ghp, Marsh:2018zyw, Chiang:2021uvt, Schive:2025bcm}, and its theoretical aspects~\cite{Chavanis:2011zi,Chavanis:2011zm,Levkov:2018kau,Guth:2014hsa,Bar-Or:2018pxz,Bar-Or:2020tys,Chan:2022bkz,Chan:2023crj,Chavanis:2019faf,Chavanis:2020upb}. The fluctuating density field, key feature that differentiates ULDM from other dark matter candidates, is expected to affect dynamical heating, friction, and relaxation of the stellar body~\cite{Amorisco:2018dcn,Church:2018sro,Lancaster:2019mde,Bar:2021jff,Marsh:2018zyw, Dalal:2022rmp, Schive:2019rrw, Yang:2024hvb, DuttaChowdhury:2023qxg}.
In particular, dynamical heating can affect the expected distributions of stars and stellar kinematics in galaxies\footnote{Notably, dark matter sub-halos also cause dynamical heating in a similar fashion as the ULDM-induced one~\cite{Penarrubia:2025auj,2026arXiv260300257P}.}. It is in fact amusing that such a simple dark matter model, without any interaction with the standard model, can have such a rich phenomenological implications. 

In our previous work~\cite{Teodori:2025rul}, we contributed to the investigation of ULDM influence on galactic scales by performing ULDM simulations for three different dwarf galaxies: Fornax, Carina, and Leo~II. That study found ULDM to be in tension with stellar kinematics and half-light radii data for $5\times 10^{-22} \lesssim m/\text{eV} \lesssim {5\times10^{-21}}$, improving upon typical constraints from Lyman-$\alpha$~\cite{Irsic:2017yje,Armengaud:2017nkf,Kobayashi:2017jcf,Leong:2018opi} and present-day stellar kinematics~\cite{Zimmermann:2024xvd}. In particular, for $m = \SI{5e-21}{\electronvolt}$, we demonstrated that dynamical heating in Leo~II and Carina is the primary driver of this tension. However, dynamical heating can be reduced by several effects~\cite{Eberhardt:2025lbx} which were not accounted for in Ref.~\cite{Teodori:2025rul}. One factor acknowledged in Ref.~\cite{Teodori:2025rul} is the neglect of stellar self-gravity. While this seems to be a fairly good approximation for these systems today, Ref.~\cite{Teodori:2025rul} pointed out that it may not have been valid at earlier times, particularly if the stellar population was denser. For Fornax, for example, it would suffice for the stellar half-light radius to have been a factor of two smaller in the past for the baryonic mass to be comparable to the dark matter mass in the central region of the galaxy. Moreover, a second caveat is the possibility of tidal stripping; a tidally stripped ULDM halo exhibits reduced dynamical heating~\cite{Chiang:2021uvt,Eberhardt:2025lbx, Yang:2025bae}. For example, Yang et al.~\cite{Yang:2025bae} recently investigated the impact of tidal stripping on Fornax for $m = \SI{e-22}{\electronvolt}$. They demonstrated that, within the current uncertainties of the Fornax orbit and the Milky Way's gravitational potential, dynamical heating may be significantly suppressed. This effect could potentially reconcile the $m = \SI{e-22}{\electronvolt}$ ULDM model with dwarf galaxy data, although constraints from the Lyman-alpha forest continue to disfavor this lower mass range~\cite{Irsic:2017yje, Armengaud:2017nkf, Kobayashi:2017jcf, Leong:2018opi}. 

In the present work, we extend previous studies by tackling both of these concerns. First, we include the full gravitational interaction between stars in our simulations, and find that stellar self-gravity can reduce the growth of the half-light radius in some cases, although not enough to qualitatively modify the main dynamical-heating conclusions of Ref.~\cite{Teodori:2025rul}. Second, we survey several dwarf galaxies using orbit integrations in different Milky Way potentials, with different orbital parameters, in order to estimate the expected degree of tidal stripping of their dark-matter halos. We then incorporate this effect in our simulations using the \textit{approximate} prescriptions of Refs.~\cite{Yavetz:2021pbc,Eberhardt:2025lbx}. Within this framework, ULDM with masses
\begin{equation}
5\times 10^{-22} \lesssim m/{\rm eV} \lesssim 5\times 10^{-21}
\end{equation}
continues to appear in tension with current dwarf-galaxy data, in agreement with the overall picture discussed in Ref.~\cite{Teodori:2025rul}. In upcoming work, we will also present results for ultra-faint dwarf galaxies~\cite{TeodoriCaputoUltra2_inprep}, which can probe higher ULDM masses; see, for example, Ref.~\cite{Dalal:2022rmp}.

The impact of stellar self-gravity and tides on stellar systems within ULDM halos, has also been considered recently in Ref.~\cite{Eberhardt:2025lbx}. That work, which we follow for the treatment of tidal stripping, however, did not attempt to model specific dwarf galaxies and compare directly with data, nor did it attempt to characterize the uncertainties associated with the tidal histories of dwarf galaxies, as we do in the present work. 

This paper is structured as follows: in Sec.~\ref{s:heating}, we describe the effects which can reduce the expected dynamical heating in a ULDM halo; in Sec.~\ref{s:sims}, we describe the setting of our simulations; we describe our results for the influence of star self-gravity and tidal stripping for Fornax/Carina/Leo II and Leo I/Draco dwarf galaxies in Sec.~\ref{s:res:ForCarLeo} and Sec.~\ref{s:draco} respectively. We discuss our results in Sec.~\ref{s:disc} and conclude in Sec.~\ref{s:conc}.

\section{Effects that can reduce ULDM dynamical heating}\label{s:heating}

In this section, we describe in more detail two effects that can reduce ULDM dynamical heating and explain how we model them in this manuscript. We first discuss tidal stripping in the MW potential, and then turn to the role of stellar self-gravity within the dwarf galaxy.

\begin{figure*}
    \centering   \includegraphics[width=0.49\textwidth]{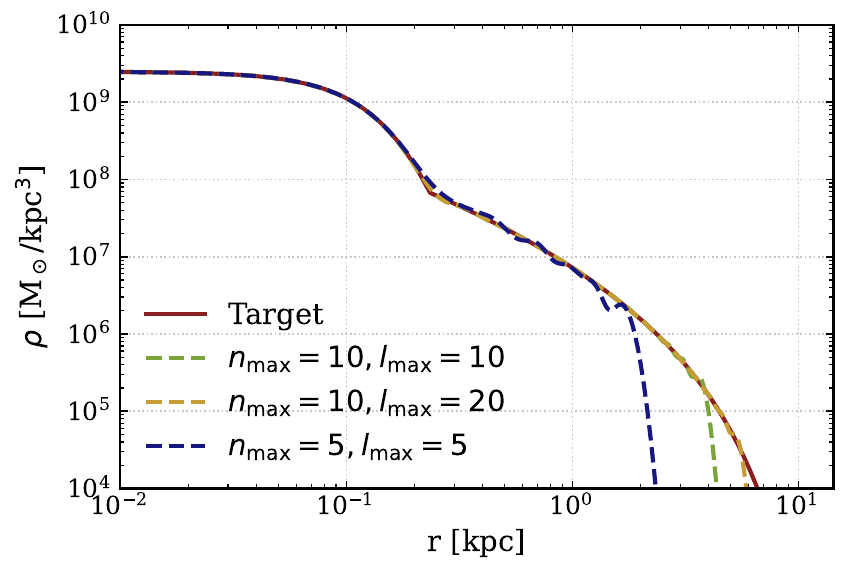}
\includegraphics[width=0.49\textwidth]{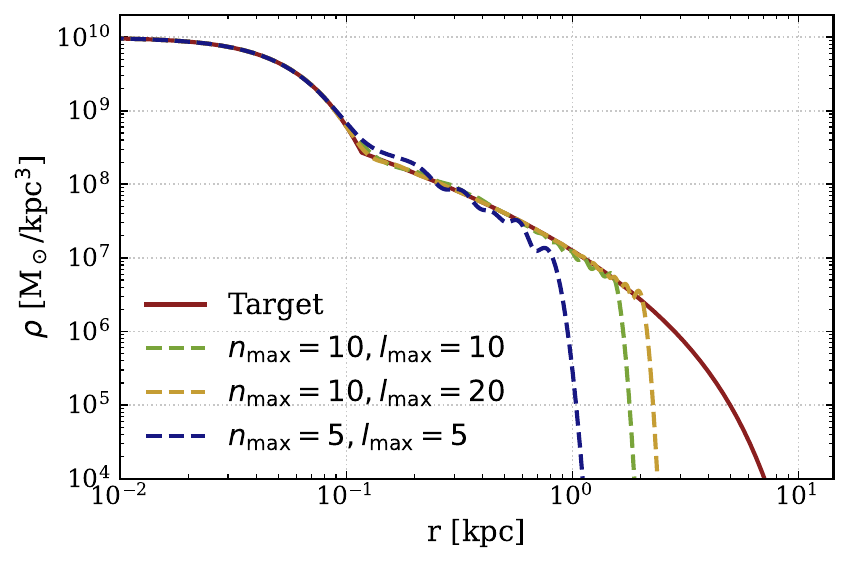}
    \caption{Two examples of halos obtained with different eigenfunction cutoffs $(n_{\rm max},l_{\rm max})$ for the same target soliton+NFW density profile, as described in Sec.~\ref{s:sims}. {\bf Left}: case with $m=\SI{1e-21}{\electronvolt}$, where the density profiles are truncated at $r_{\rm t}\simeq (2,4,6)\,\mathrm{kpc}$ for different cutoff choices. {\bf Right}: case with $m=\SI{2e-21}{\electronvolt}$, where the density profiles are truncated at $r_{\rm t}\simeq (1,2,2.5)\,\mathrm{kpc}$.}
    \label{fig:tidal_example}
\end{figure*}

\subsection{Tidal stripping of the external halo}
Following Ref.~\cite{Yavetz:2021pbc}, the ULDM field can be heuristically decomposed in energy eigenfunctions $\psi_n$ with eigenvalue $E_n$ as 
\begin{equation} \label{eq:simple_expansion}
\psi \sim \sum^{N_{\rm max}}_{n=0} \psi_n \e^{\iu \varphi_n - \iu E_n t} \ ,
\end{equation}
where $\varphi_n$ are random phases, and $N_{\rm max}+1$ is the total number of eigenfunctions, ordered by the value of the corresponding $E_n$. The density field then reads
\begin{align}
\begin{aligned}
\rho &= |\psi|^2 \sim \sum_{n=0}^{N_{\rm max}}|\psi_n|^2 \\
&+ \sum_{n\neq m}^{N_{\rm max}} \psi_n\psi^*_m \e^{\iu (\varphi_n - \varphi_m) - \iu (E_n - E_m)t} \ .
\end{aligned}
\end{align}
In particular, interference effects, of which dynamical heating is a notable example, may be seen to come from the $n\neq m$ terms. 

Tidal stripping can remove mass in the outer parts of the halo. As higher $n$ eigenfunctions have support for larger radii, in practice a tidally stripped halo has a reduced number of eigenfunctions, i.e. a cutoff on $N_{\rm max}$. As the fluctuating density field comes from interference terms between different $n$ eigenfunctions, removal of high $n$ eigenfunctions is expected to reduce the strength of such fluctuations, and hence reduce dynamical heating. 

A possible way to implement a tidally stripped halo comes from the eigenfunction method of Ref.~\cite{Yavetz:2021pbc}, discussed in this particular context on Ref.~\cite{Eberhardt:2025lbx}. The method is summarized in App.~\ref{s:eigen}. The resulting $\psi$ decomposition, Eq.~\eqref{eq:sum_eigen}, is very similar to the heuristic Eq.~\eqref{eq:simple_expansion}, with the difference that the cutoff on the eigenfunctions sum is labeled by two numbers $(n_{\rm max},l_{\rm max})$. We remark that, in our simulations, we use such eigenfunction decomposition only to inform the initial conditions of the ULDM halo. The ULDM evolution uses the full Schr\"odinger-Poisson solver, detailed in App.~\ref{s:code}.

Within the eigenfunction construction described in App.~\ref{s:eigen}, a tidally stripped halo with tidal radius $r_{\rm t}$ can be modeled by truncating the set of populated modes such that the reconstructed density profile falls rapidly beyond $r_{\rm t}$ (equivalently, by fitting a target profile that is truncated at $r_{\rm t}$ and retaining only modes up to $(n_{\rm max},l_{\rm max})$ needed to reproduce it). Since the time-dependent density fluctuations arise from interference among populated modes, reducing the number of contributing modes, in particular the more weakly bound, spatially extended ones, suppresses the amplitude of these fluctuations and can therefore reduce dynamical heating. In Fig.~\ref{fig:tidal_example} we show examples of the same halo reconstructed with different $(n_{\rm max},l_{\rm max})$. We remark that the $(n_{\rm max},l_{\rm max})$ are chosen such that the resulting density profile plunges to zero around the target $r_{\rm t}$, see the right panel of Fig.~\ref{fig:LeoII_5em21_tides}. There may be slightly different $(n_{\rm max},l_{\rm max})$ values which give similar $r_{\rm t}$, although this latter depends mainly on $n_{\rm max}$. We verified that initializing the halo with different $(n_{\rm max},l_{\rm max})$ which maintain a similar $r_{\rm t}$ do not affect results. For sufficiently small cutoffs, the resulting dynamical heating of stars can be significantly reduced; see Ref.~\cite{Eberhardt:2025lbx}. 

Implementing the effect of tidal stripping as a hard cutoff in $(n_{\max},l_{\max})$ is a very simplistic approach and far from what happens in reality; however, we believe that such a procedure yields a conservative (i.e., minimal-heating) estimate compared to more realistic tidal evolution~\cite{Yang:2025bae}, which is time-dependent and may retain residual power in weakly bound or unbound components. Where a comparison is possible, our results and Ref.~\cite{Yang:2025bae} indeed appear consistent with this expectation.

In practice, to set the initial conditions of our simulations, we must assign an \textit{expected} tidal radius $r_{\rm t}$ to each dwarf galaxy. To this end, we reconstruct the tidal histories of the dwarfs in different Milky Way potentials, while propagating the observational uncertainties on their orbits. More specifically, for each Monte Carlo realization of the distance, line-of-sight velocity, and proper motions, we compute the minimum value of the tidal radius reached along the orbit, normalized to the half-light radius,
$\min_t \!\left[\frac{r_{\rm t}(t)}{r_{1/2}}\right]$. For each dwarf, Tab.~\ref{tab:dwarf_inputs} reports the median of this quantity together with its 2.5th--97.5th percentile interval over the Monte Carlo sample. In our simulations, the quantity used to define the stripped initial conditions is the 2.5th percentile of $\min_t \!\left[\frac{r_{\rm t}(t)}{r_{1/2}}\right] \,$, that is, a conservative estimate of the smallest tidal radius compatible with the orbital uncertainties. A detailed description of the tidal-history reconstruction is given in App.~\ref{s:orbits}; here we limit ourselves to summarizing the orbital properties and tidal diagnostics of the dwarf sample considered in this work in Tab.~\ref{tab:dwarf_inputs}.

\begin{table*}[t]
\centering
\caption{Astrometric and kinematic inputs for orbit integrations. Positions are in Galactic longitude $l$ and latitude $b$. For consistency we adopt systemic proper motions from Ref.~\cite{Battaglia_2022} (see their Table~B.2). Distances $D_\odot$ and systemic heliocentric line-of-sight velocities $v_{\rm los}$ are kept as in the compilation used in the main text. The last three columns give the fiducial halo mass $M_{200}$ and NFW concentration $c$~\cite{Read:2018fxs}, the half light radius~\cite{McConnachie2012, Walker:2009zp, Koch:2007ye} and the tidal-to-half-light ratio $\min_t(r_{\rm t}/r_{1/2})$ quoted as the median with errors spanning the $2.5$th to $97.5$th percentiles of the Monte Carlo orbit ensemble.}
\label{tab:dwarf_inputs}
\setlength{\tabcolsep}{5.2pt}
\renewcommand{\arraystretch}{1.15}
\footnotesize
\begin{tabular}{lccccccccc}
\hline
Name & $l$ [deg] & $b$ [deg] & $D_\odot$ [kpc] & $v_{\rm los}$ [km s$^{-1}$] & $\mu_{\alpha*}$ [mas yr$^{-1}$] & $\mu_{\delta}$ [mas yr$^{-1}$] & $M_{200}\,(c)$ [$M_\odot$] & $r_{1/2} [\rm kpc]$ & $r_{\rm t}/r_{1/2}$ \\
\hline
Fornax & 237.1 & $-65.7$ & $147 \pm 12$ & $55.3 \pm 0.1$   & $0.381 \pm 0.001$ & $-0.358 \pm 0.002$ & $10^{10}\,(11)$ & $0.668 \pm 0.034$ & $8.5^{+2.9}_{-2.4}$ \\
Carina & 260.1 & $-22.2$ & $105 \pm 6$  & $222.9 \pm 0.1$  & $0.53 \pm 0.01$    & $0.13 \pm 0.01$     & $0.8 \times 10^9\,(15)$ & $0.24 \pm 0.02$ & $23.5^{+2.2}_{-11.1}$ \\
Leo II  & 220.2 & $+67.2$ & $233 \pm 14$ & $78.3 \pm 0.6$   & $-0.11 \pm 0.03$ & $-0.14 \pm 0.03$    & $10^9\,(20)$ & $0.21 \pm 0.02$ & $13.7^{+33.6}_{-12.2}$ \\
Leo I  & 226.0 & $+49.1$ & $254 \pm 15$ & $282.5 \pm 0.1$  & $-0.05 \pm 0.01$   & $-0.11 \pm 0.01$    & $5.6 \times 10^{9}\,(11)$ & $0.25 \pm 0.03$ & $10.1^{+7.6}_{-6.3}$ \\
Draco  & 86.4  & $+34.7$ & $76 \pm 6$   & $-292.3 \pm 0.4$ & $0.04 \pm 0.01$  & $-0.19 \pm 0.01$    & $1.8\times10^9\,(15)$ & $0.22 \pm 0.02$ & $12.2^{+3.6}_{-3.1}$ \\
\hline
\end{tabular}
\end{table*}

\subsection{Stellar self-gravity}

Another effect which can reduce dynamical heating is star self gravity. If the stellar component is compact enough such that its potential dominates (or it is not too far from) the ULDM one, then stars are less affected by the ULDM potential fluctuations. In this case, of course, the ULDM fluctuations themselves do not change (unlike in the presence of tidal stripping). However, a deeper background potential from the stellar component implies a smaller fractional change in orbital energy per kick and typically a shorter dynamical time. This can make the stellar system more ``adiabatic'' with respect to the ULDM fluctuation field, thereby reducing the efficiency of diffusion.

To assess when it is justified to neglect stellar self-gravity, it is useful to distinguish a few commonly used (and sometimes conflated) quantities.
Within a given radius $r$, we define the stellar mass $M_\star(r)$, the dark-matter mass $M_{\rm DM}(r)$, and the luminosity $L(r)$ (typically in the $V$ band).
The \emph{total} mass-to-light ratio is\footnote{Mass-to-light ratios are band dependent. Throughout, unless stated otherwise, we quote $M/L_V$ (i.e.\ using the $V$-band luminosity $L_V$), as commonly adopted in the dSph literature (e.g.\ Ref.~\cite{lokas_2009}). When converting to a stellar mass, we correspondingly use the stellar mass-to-light ratio $\Upsilon_{\star,V}\equiv M_\star/L_V$.}
\begin{equation}
\left(\frac{M}{L}\right)(r)\equiv \frac{M_{\rm DM}(r)+M_\star(r)+M_{\rm gas}(r)}{L(r)}\,,
\end{equation}
while the \emph{stellar} mass-to-light ratio is
\begin{equation}
\Upsilon_\star(r)\equiv \left(\frac{M_\star}{L}\right)(r)\,.
\end{equation}
These two quantities are related to the dark-to-stellar mass ratio via
\begin{equation}
\frac{M_{\rm DM}}{M_\star}(r)=\frac{(M/L)(r)}{\Upsilon_\star(r)}-1-\frac{M_{\rm gas}(r)}{M_\star(r)}\simeq \frac{(M/L)(r)}{\Upsilon_\star(r)}-1\,,
\label{eq:MDM_over_Mstar}
\end{equation}
where the last step uses the fact that classical dSphs are extremely gas-poor: upper limits on the neutral hydrogen content within the half-light ellipses imply $M_{\rm HI}/L_V\sim 10^{-3}\,M_\odot/L_\odot$ for the Galactic dSph population, i.e.\ negligible compared to the stellar contribution \cite{Spekkens_2014}. With these definitions in mind, the common statement that ``dSphs are dark-matter dominated'' is correct, but it does not automatically imply that stellar self-gravity is irrelevant. What matters for the stellar dynamics is the \emph{enclosed} ratio $M_{\rm DM}(r)/M_\star(r)$ in the region probed by the stars, and this ratio can be only moderately large in some systems.
For Fornax, Ref.~\cite{Battaglia_2015} finds that the dark-to-luminous mass ratio is $M_{\rm DM}/M_{\rm lum}\simeq 5$--$6$ within $r\simeq 1.6~{\rm kpc}$ (and $16$--$18$ within $3~{\rm kpc}$), implying that even today the stellar component can contribute non-negligibly to the central gravitational potential.
By contrast, Carina and Leo~II appear more dark-matter dominated.
For Carina, dynamical modelling yields a total mass-to-light ratio $M/L\simeq 66$ (with sizeable uncertainties)~\cite{lokas_2009}, whereas for Leo~II Ref.~\cite{Koch:2007ye} reports a global $M/L\simeq 27$--$45$.
Adopting the typical range $\Upsilon_\star\sim\mathcal{O}(1$--$3)$ expected for old, metal-poor stellar populations (e.g.\ $\Upsilon_\star\simeq 1.8\pm 0.8$ for Carina from its resolved star-formation history \cite{de_Boer_2014}), Eq.~\eqref{eq:MDM_over_Mstar} implies that $M_{\rm DM}/M_\star$ is generally of order $\sim 20$--$60$ for Carina (for $M/L\simeq 66$) and of order $\sim 8$--$40$ for Leo~II (for $M/L\simeq 27$--$45$), depending on the assumed $\Upsilon_\star$.Thus, even in these more dark-matter dominated dwarfs, the hierarchy between the dark and stellar components is not parametrically large. This point is particularly relevant because the importance of stellar self-gravity is sensitive not only to the \emph{total} stellar mass, but also to its \emph{concentration}.
For fixed $M_\star$, making the stellar distribution more compact increases the enclosed $M_\star(r)$ and hence the stellar contribution to the background potential, $\Phi_\star(r)\sim G M_\star(r)/r$. Therefore, while neglecting stellar self-gravity may be a reasonable approximation for present-day dwarfs in many cases, it need not have been valid at earlier times if the stellar component was denser. This motivates treating stellar self-gravity explicitly when assessing ULDM-induced dynamical heating, and in particular when revisiting the heating-driven tension identified in Ref.~\cite{Teodori:2025rul}. In practice, this simply amounts in inserting the massive stars potential in the ULDM equations of motion as an external potential, see App.~\ref{s:code}.

Finally, even in systems where ULDM dominates the star component at all times, dynamical heating may be significantly reduced if the mass of a single star is not much smaller than the mass associated to ULDM fluctuations. In fact, from the results of Ref.~\cite{Bar-Or:2018pxz,Bar-Or:2020tys}, ULDM can effectively be seen to be composed by quasi-particles of mass\footnote{We remark that such picture is accurate only in the homogeneous ULDM background approximation. Its applicability to real systems, which also include soliton fluctuations which cannot be modelled via this quasi-particle picture, cannot be trusted more than an order of magnitude estimate.}
\begin{align}
\begin{aligned}
M_{\rm QP} \sim& \SI{2e7}{M_\odot} \qty(\frac{\rho}{\SI{0.01}{M_\odot\per\parsec\cubed}})\\
&\times\qty(\frac{\SI{e-21}{\electronvolt}}{m})^3 \qty(\frac{\SI{10}{\kilo\meter\per\second}}{\sigma} )^3 \ .
\end{aligned}
\end{align}
For the benchmark values we consider in this work, $M_{\rm QP}$ will always be much greater than the typical mass of the star. Nevertheless, our implementation of stars dynamics, to be described in App.~\ref{s:code}, takes into account this effect as well, since backreaction of stars on the ULDM field is explicitly present in the solver, while in many other cases this is not properly taken into account. For example, the limits shown in Ref.~\cite{May:2025ppj} using UNIONS~I for $m\sim \SI{e-17}{\electronvolt}$ are not exempt from this caveat\footnote{We thank K.~Blum for an assessment of this issue.}. For such an object and ULDM mass, one finds
\begin{align}
\begin{aligned}
M_{\rm QP} \sim& \SI{e-3}{M_\odot} \qty(\frac{\rho}{\SI{0.1}{M_\odot\per\parsec\cubed}})\\
&\times\qty(\frac{\SI{e-17}{\electronvolt}}{m})^3 \qty(\frac{\SI{5}{\kilo\meter\per\second}}{\sigma} )^3 \, .
\end{aligned}
\end{align}
This is much smaller than a typical stellar mass. However, Ref.~\cite{May:2025ppj} follows Ref.~\cite{Dalal:2022rmp} and models the stellar component as \textit{massless test particles} evolved in the mean plus fluctuating FDM potential. This corresponds to neglecting both stellar self-gravity and any backreaction of the ULDM field induced by the stellar component. In the regime $M_{\rm QP}\ll M_\star$, such backreaction effects may not be negligible, and the passive-tracer approximation does not apply. We leave a proper treatment of this regime for future work, although we stress that the nature of UNIONS~I is unclear and it may not be dark matter dominated. In fact, soon after Ref.~\cite{May:2025ppj} appeared, Ref.~\cite{cerny2025observationalevidencedarkmatter} presented new Keck spectroscopy of additional members and found a 95\% CL upper limit $\sigma_v<2.3~{\rm km\,s^{-1}}$, with the data strongly favoring a stellar-only dispersion $\sigma_\star\simeq 0.1~{\rm km\,s^{-1}}$, concluding that there is currently no observational evidence for dark matter in the system. Chemically, the same work finds a small metallicity-spread limit, and argues that the overall properties are more consistent with a star-cluster interpretation.

\section{Simulation setting} \label{s:sims}
We now describe the ULDM simulations we performed, tackling the dynamical heating caveats described in Sec.~\ref{s:heating}. 
The simulations we performed where tidal stripping is not taken into account are equivalent to the ones shown in Ref.~\cite{Teodori:2025rul}. We refer to that and App.~\ref{s:code} for details. In particular, such simulations initialize the ULDM halo using the Eddington procedure outlined in the Appendix A of Ref.~\cite{Teodori:2025rul}. The only difference with the simulations of Ref.~\cite{Teodori:2025rul} is the implementation of stellar mutual interactions and feedback on the ULDM field, in the way described in App.~\ref{s:code}.

For ULDM halos where tidal stripping is taken explicitly into account, we instead follow the procedure of Ref.~\cite{Yavetz:2021pbc}, which we outlined in Sec.~\ref{s:heating}.
In particular, 
the target profile of the eigenmode solver can be taken to be composed of a soliton profile $\rho_{\rm c}$  (which is the ground state solution of the Schr\"odinger-Poisson equations) and a Navarro-Frenk-White (NFW)~\cite{Navarro:1996gj} profile $\rho_{\rm NFW}(r)$, separated by a transition radius $r_{\rm r}$. Explicitly,
\begin{align}
\begin{aligned}
\rho(r) = \begin{cases}
&\rho_{\rm c} (r) \eta_{\rm t}(r) \ , \ r \le r_{\rm r} \ , \\
&\rho_{\rm NFW}(r) \eta_{\rm t}(r) \ , \ r > r_{\rm r} \ ,
\end{cases}
\end{aligned}
\end{align}
where we added a tidal envelope $\eta_{\rm t}$, which suppresses the density after the tidal radius $r_{\rm t}$,
\begin{equation}
\eta_{\rm t}(r) = \e^{-r^2/(2 r^2_{\rm t})} \ .
\end{equation}
We will later comment on the choice of the tidal radius, but we here note that we impose a cutoff on the energy eigenvalues such that the resulting reconstructed density profile drops after $r_{\rm t}$, see Fig.~\ref{fig:tidal_example} for examples.

The soliton profile can be parametrized as~\cite{Schive:2014dra}
\begin{equation}
\rho_{\rm c} (r) = \frac{1.9\times 10^{7}\, (m/10^{-22} \mathrm{eV})^{-2} (r_{\rm c}/\mathrm{kpc})^{-4}}{\qty(1+0.091 \qty(r/r_{\rm c})^2)^{8}}  M_\odot/\mathrm{kpc}^3 \ ,
\end{equation}
where $r_{\rm c}$ is the core radius, which we can consider as a free parameter. In this work, we take $r_{\rm r} = 2.5 r_{\rm c}$, which is roughly the transition radius found in simulations. 
$\rho_{\rm NFW}$ instead reads as
\begin{equation}
\rho_{\rm NFW} = \frac{\rho_{\rm s} r^3_{\rm s}}{r(r+r_{\rm s})^2} \ ;
\end{equation}
$r_{\rm s}$ is the NFW scale radius, free parameter, whereas $\rho_{\rm s}$ is determined by matching the densities at $r_{\rm r}$.
Once initialized, we let the halo relax for $\sim 1$ Gyr before inserting stars, to wait out for artificial initial condition transients.
This eigenstate initialization method could also be used for simulations where tides are not taken into account; one would just need to insert enough eigenstates to fill the box. We still stick with the Eddington procedure for the no-tidal stripping case here following Ref.~\cite{Teodori:2025rul}, but we notice in Sec.~\ref{s:res:ForCarLeo} that, when $r_{\rm t}$ is high enough, we find no difference between Eddington initialization and eigenstate initialization.

\section{Results for Fornax, Carina and Leo II}
\label{s:res:ForCarLeo}
In this section we study the three systems simulated in Ref.~\cite{Teodori:2025rul}, investigating the impact of stars self-gravity and tidal stripping. For stars self-gravity, which has its main effect on dynamical heating, we only show results for Carina and Leo II, whose limits are driven by dynamical heating. For tidal stripping, we study all three systems.

\subsection{Results including stars self-gravity} \label{s:self_LeoII}

Unlike Fornax, the ULDM tension in Leo II and Carina is driven by dynamical heating as mentioned above. It is therefore important to assess how sensitive this effect is to stellar self-gravity in Leo II- and Carina-like systems, although Carina and Leo II appear to have substantially larger dark-to-luminous mass ratios compared to a system like Fornax. To this end, we perform simulations following the same general procedure adopted here and in Ref.~\cite{Teodori:2025rul}, taking $m=\SI{5e-21}{\electronvolt}$ and fixing the initial ULDM-to-stellar mass ratio to be $\sim 10$ within $\SI{0.7}{\kilo\parsec}$. In particular, we simulate $10^4$ stars, each with mass $100\,M_\odot$.

\begin{figure}
    \centering
    \includegraphics[width=0.49\textwidth]{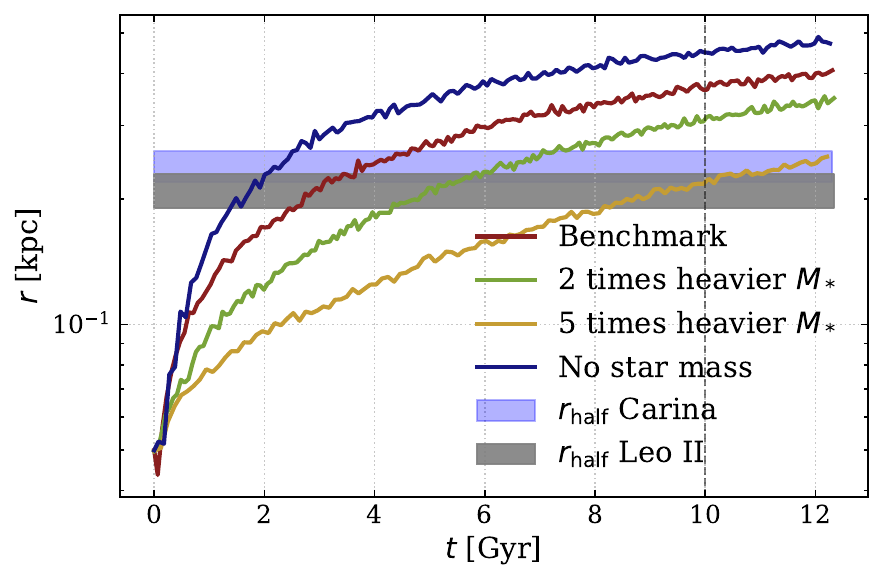}
    \caption{Examples of a simulation run, targeting Carina/Leo II-like systems, for $m=\SI{5e-21}{\electronvolt}$ with star self-gravity, showing the $r_{\rm half}$ evolution over time, compared with the observed one. We show the results of four runs, each one of them with a different mass of the stars. The benchmark is what it is expected for Leo II, $M_* \sim 10^6 M_\odot$, while for Carina everything should be in fact reduced by a factor of $\sim 2$.
    }
    \label{fig:leoII}
\end{figure}

Results are presented in Fig.~\ref{fig:leoII}, where we initialize the simulations with a Plummer half-light radius of $0.05$~kpc, i.e. about four to five times smaller than the present-day observed value, while keeping the total number of stars fixed. For a Plummer profile, this corresponds to a central stellar density 
\begin{equation}
\rho_{*,0}=\frac{3M_*}{4\pi a^3}\sim 2 \,\qty(\frac{M_*}{10^6 M_\odot}) \qty(\frac{\SI{50}{\parsec}}{a})^3 \,M_\odot\,{\rm pc}^{-3} \ ,
\end{equation}
for the present-day stellar masses of Leo~II/Carina. This is roughly one to two orders of magnitude denser than all present-day classical dwarf spheroidals, and also above the average stellar density in globular clusters (but several orders of magnitude below central cores densities).

\begin{figure*}
\centering \includegraphics[width=0.49\linewidth]{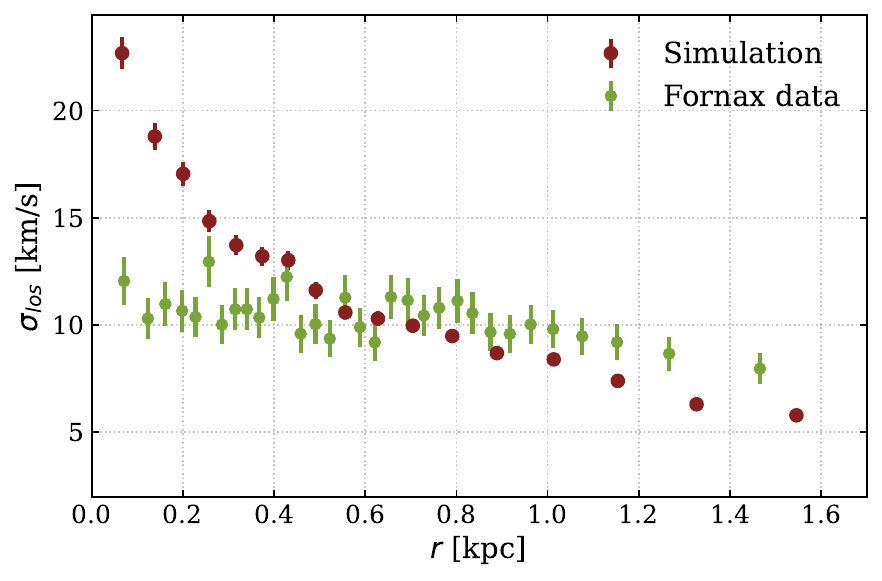} \includegraphics[width=0.49\linewidth]{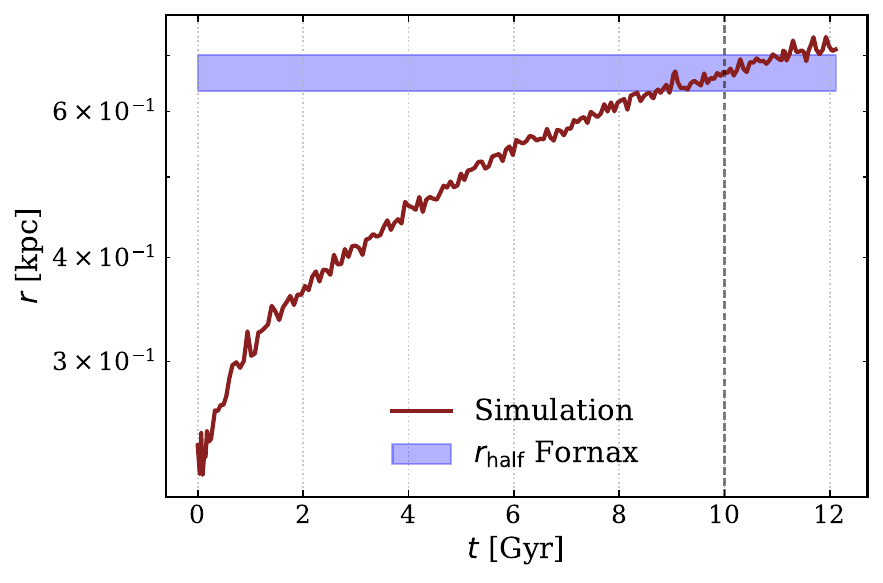} \caption{Fornax simulation, with tides, for $m=\SI{e-21}{\electronvolt}$. Dynamical heating is slowed down with respect to the examples of Ref.~\cite{Teodori:2025rul}, but the inner peak on the dispersion curve is retained. } \label{fig:Fornax_1em21} 
\end{figure*}

\begin{figure*}
\centering \includegraphics[width=0.49\linewidth]{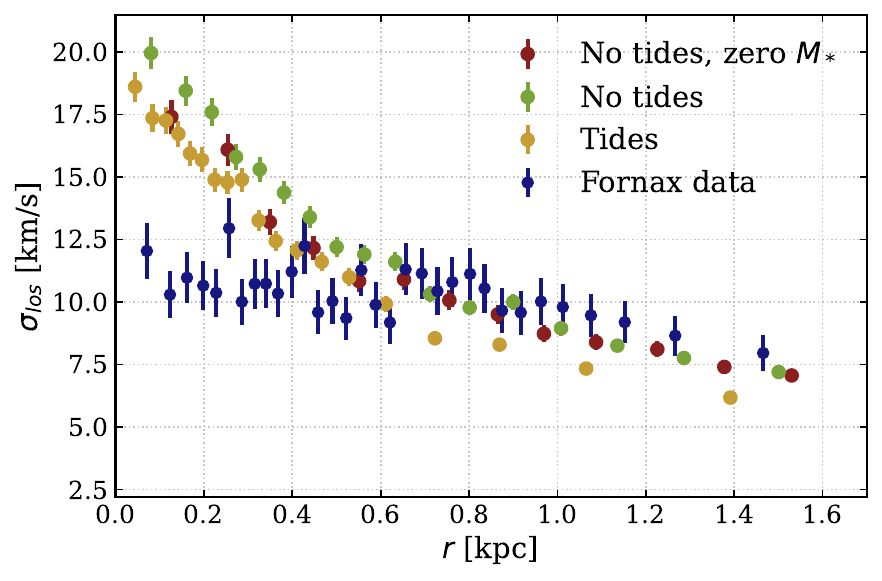} \includegraphics[width=0.49\linewidth]{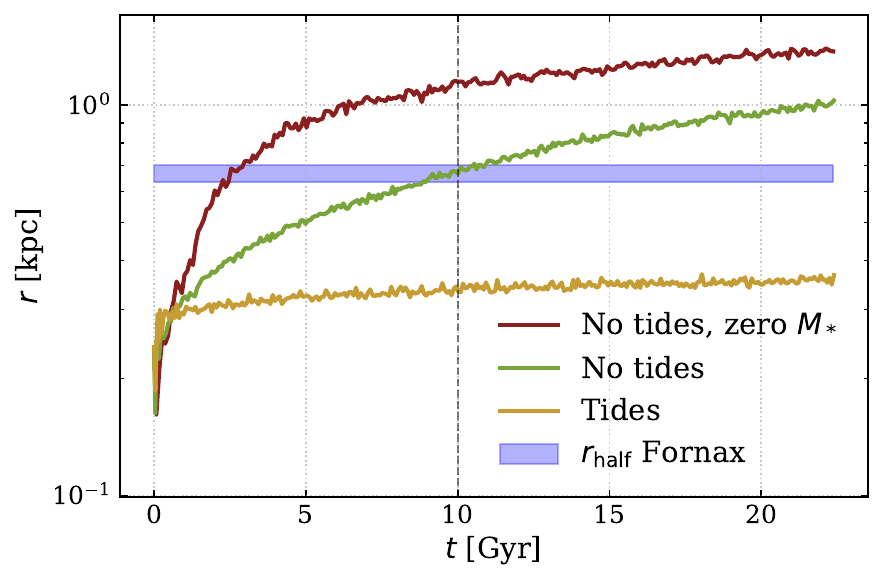} \caption{Fornax simulations, for $m=\SI{5e-22}{\electronvolt}$. We compare three runs, one with tides, one without and one both without tides and with zero $M_*$. Although dynamical heating is basically stopped for the tide case, the inner peak on the velocity dispersion curve due to the soliton would put this mass at odds with data nevertheless. Notice also how the non-zero mass of the stars affects the growth of $r_{\rm half}$.} \label{fig:5em22_Fornax_tides} 
\end{figure*}

Although, as expected, the growth of the half-light radius is somewhat slower than in the examples shown in Ref.~\cite{Teodori:2025rul}, it remains large enough to make ULDM with \(m=\SI{5e-21}{\electronvolt}\) incompatible with the data. We stress that tidal stripping is not included in these simulations, but it will be considered, together with star self-gravity, in Sec.~\ref{s:res_tidal}. To illustrate the sensitivity of the results to stellar self-gravity, Fig.~\ref{fig:leoII} also shows examples of the half-light-radius evolution for Leo II/Carina simulations with the same ULDM halo parameters, but with stars that are 2 and 5 times more massive (plus a no mass case for comparison). In the 5 times more massive case, this corresponds to \(M_{\rm DM}/M_* \sim 2\), i.e. to a mass-to-light ratio similar to that of Fornax, and a stellar density at $0.1$ pc roughly twice the dark matter one. As expected, the growth of the half-light radius becomes progressively slower as the stellar mass increases. We note that it is conceivable, and perhaps even expected, that the stellar component may have been larger by a factor of \(\sim 2\) in the past, simply because stars return part of their mass to the interstellar medium during their evolution~\cite{Bruzual:2003tq}. Reaching instead a stellar-to-dark-matter mass ratio larger by a factor of 5 would require extreme tidal effects together with a cuspy central dark matter profile~\cite{Penarrubia:2007zx}. In such a case, the dark matter cusp would be more resilient to tides than the stellar core, potentially leading to systems that are more dark-matter dominated in their centers today than they were 10~Gyr ago. This is not expected to occur, however, if the dark matter profile is also cored, as in fuzzy dark matter. Finally, we emphasize that even in cases where the half-light radius changes substantially over the lifetime of the dwarf, the fraction of stars effectively ``lost'' due to dark matter heating never exceeds the percent level within our simulations. The case with a stellar mass larger by a factor of 5 should therefore be regarded as an extreme scenario, shown mainly for reference. Of course, one could imagine changing the stellar concentration rather than total mass, for example reducing the initial half-light radius even further for systems like Carina or Leo II. However, we have already considered the initial half-light radius to be 5 times smaller in the past, which implies a density about $\sim 120$ times larger. Reducing the half-light radius even further would lead to globular-cluster-like densities. In other words, we would be considering entirely different progenitors, not simply the same objects in a denser configuration.

\begin{figure*}
    \centering
    \includegraphics[width=0.475\linewidth]{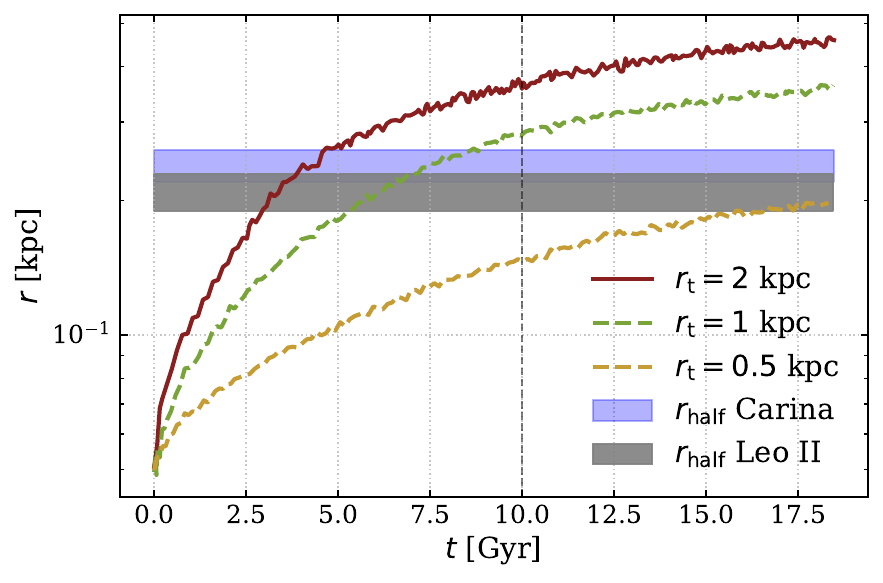}
    \includegraphics[width=0.49\linewidth]{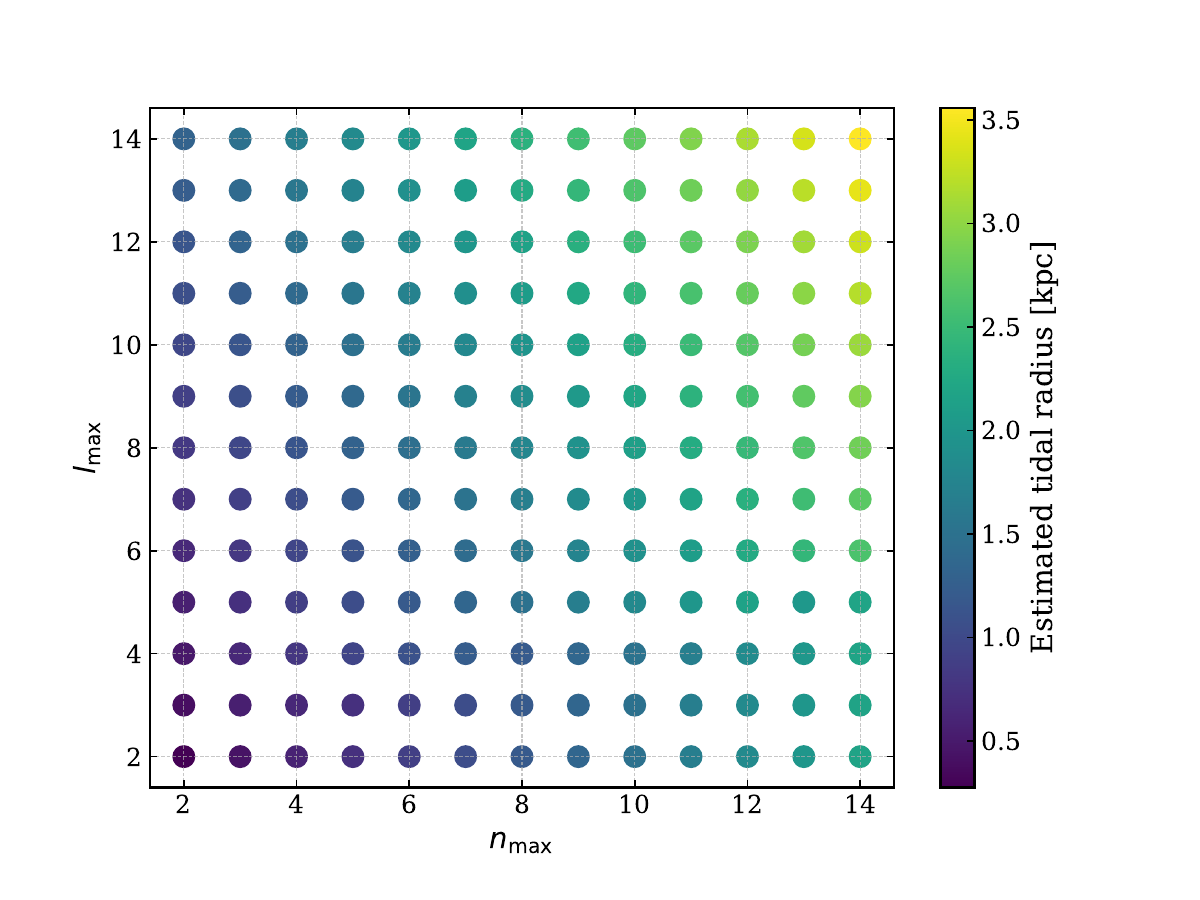}
    \caption{Leo II/Carina simulation, with tides ($r_{\rm t} = 0.5$ kpc, $r_{\rm t} = 1$ kpc, and $r_{\rm t} = 2$ kpc), for $m=\SI{5e-21}{\electronvolt}$. 
    {\bf Left}: $r_{\rm half}$ as a function of time. {\bf Right}: Estimated tidal radius for a target profile, as a function of $(n_{\rm max}, l_{\rm max})$. For definiteness, the tidal radius is estimated as the radius where the density drops by a factor of 100 within a 0.2 kpc radial window. Notice that the $n_{\rm max}$ dependence is stronger than the $l_{\rm max}$ dependence.}
    \label{fig:LeoII_5em21_tides}
\end{figure*}

\subsection{Results including also tidal stripping} \label{s:res_tidal}
We now describe our results including the impact of tidal stripping, together with stars self-gravity. Fornax, Carina and Leo II are possibly affected by tides in a different extent, as can be seen in Tab.~\ref{tab:dwarf_inputs}. Then, implementation of some form of tidal stripping is required in order to not have a biased estimate of the amount of dynamical heating. Coming back to the results of Ref.~\cite{Teodori:2025rul}, the constraint for $m=\SI{e-21}{\electronvolt}$ in Fornax was not based on dynamical heating, but rather was due to the inner velocity peak on the velocity dispersion curve. Such a peak was caused by the growing soliton. A possible question is whether a tidally stripped halo would form a soliton within Fornax which is massive enough to give rise to a prominent central peak, not seen in data. 

A proper assessment of the problem should go towards understanding how solitons form during collapse of the halos originating the dwarf galaxies we see. In general, collapse and merging of galaxies would form solitons on scales which are much faster than the relaxation time-scales, the latter being a better proxy for our simulated halos. However, what we can do is to initialize a simulation with a tidally stripped halo, whose initial soliton size, together with the tidally stripped NFW tail, is such that  its induced dispersion velocity curve would still be not too incompatible with data, and study its evolution.

When considering tidally stripped halos, we take as $r_{\rm t}$ the lower 2-$\sigma$ (which roughly corresponds to the $2.5$ percentile of the $r_{\rm t}$ distribution found via Monte Carlo sampling) value from Tab.~\ref{tab:dwarf_inputs}, with the sole exception of Leo II. In particular, for Fornax we take $r_{\rm t} \simeq \SI{3}{\kilo\parsec}$, and $r_{\rm t} \simeq \SI{2}{\kilo\parsec}$ for Carina. For Leo II, this criterion would give a halo which is entirely tidally stripped beyond the stellar body. This happens because Leo II is relatively far, and the determination of the orbit is unavoidably worse, allowing in extreme cases for very close encounters with the Milky Way. This can very well be what actually happens, but given the uncertainties involved, we show results for Leo II where the tidal radius is $0.5$ kpc, $1$ kpc and $2 $ kpc (the latter like Carina).

With the possible exception of Leo~II, this choice should make our estimate of the appropriate tidal radius for each system fairly conservative, especially given that, among the Milky Way potentials we considered, we adopted the most conservative one, namely the one yielding the smallest pericenters. As we discuss further in Sec.~\ref{s:disc}, in comparison with Ref.~\cite{Yang:2025bae}, we also have reasons to believe that our very method of implementing tidal stripping is conservative. In fact, we remind the reader that we impose tidal stripping as a sharp truncation of the initial halo through a cutoff in the \((n,\ell)\) eigenstate expansion. This seems a more aggressive prescription than explicitly evolving the halo within the Milky Way potential, as done in Ref.~\cite{Yang:2025bae}, and should therefore, if anything, tend to overestimate rather than underestimate the impact of tides.

\begin{figure*}
    \centering
    \includegraphics[width=0.49\linewidth]{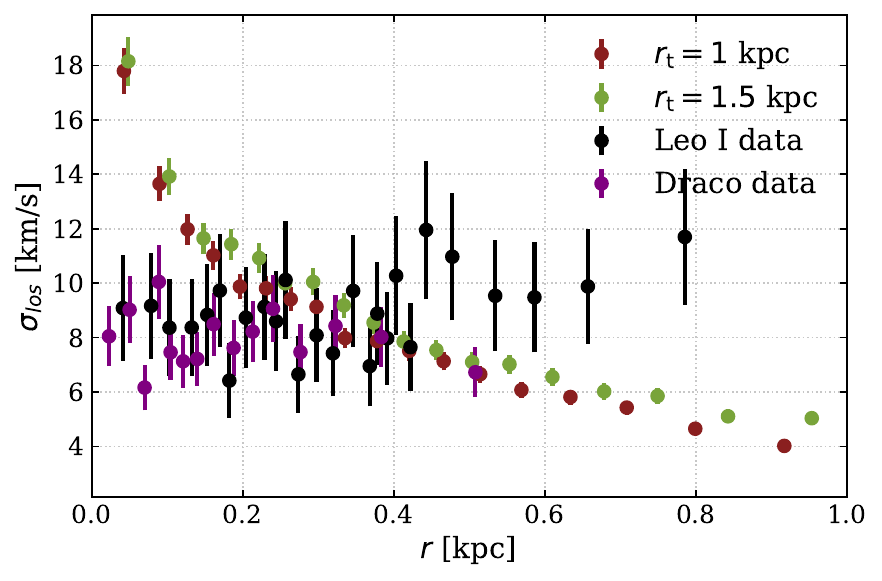}
    \includegraphics[width=0.49\linewidth]{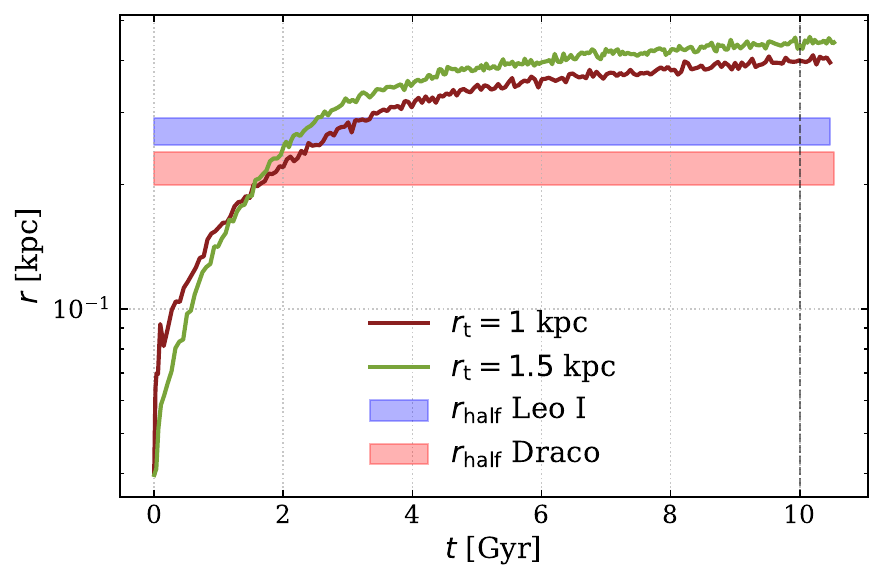}
    \caption{Run for $m=\SI{2e-21}{\electronvolt}$ for Leo I/Draco. Both stellar kinematics and dynamical heating make data incompatible with ULDM of this mass.}
    \label{fig:LeoI_2em21}
\end{figure*}

\begin{figure*}
    \centering
    \includegraphics[width=0.49\linewidth]{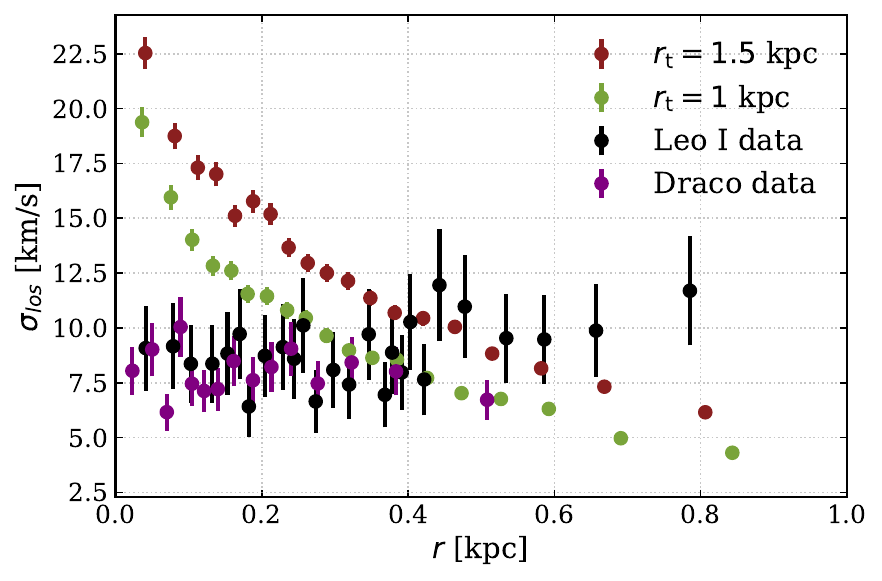}
    \includegraphics[width=0.49\linewidth]{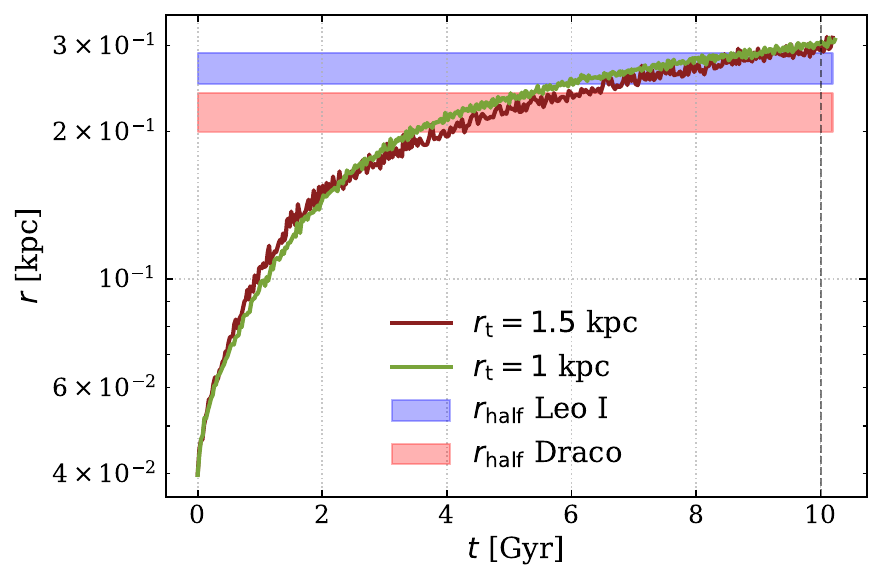}
    \caption{Runs for $m=\SI{5e-21}{\electronvolt}$ for Leo I/Draco. Mostly stellar kinematics make data incompatible with ULDM of this mass. Notice that heating is mostly unaffected by the different $r_{\rm t}$ considered in this case. }
    \label{fig:LeoI_5em21}
\end{figure*}

We show such an exercise for Fornax in Fig.~\ref{fig:Fornax_1em21} and Fig.~\ref{fig:5em22_Fornax_tides}, for two dark matter masses, $m = 10^{-21}, \, 5\times 10^{-22}$ eV. Both cases remain in tension with the data because of the inner peak in the stellar kinematics, irrespective of tides. Notice, however, that dynamical heating is suppressed when tides are taken into account, in qualitative agreement with what was also pointed out for Fornax in Ref.~\cite{Yang:2025bae}. This is also consistent with the results of Ref.~\cite{Teodori:2025rul}, which had already anticipated that, for these masses, the slowing down of heating due to tides would not invalidate the constraints, since they are driven mainly by the peaks in the velocity dispersion.

Then, in Fig.~\ref{fig:LeoII_5em21_tides}, we show an example of a Leo~II/Carina simulation for $m=\SI{5e-21}{\electronvolt}$ with tidal stripping included. The results for $r_{\rm t}=2$~kpc, which corresponds to the lower 2.5th-percentile value inferred for Carina from the tidal-to-half-light-radius ratio $\min_t(r_{\rm t}/r_{1/2})$ obtained from our \texttt{galpy} simulations described in App.~\ref{s:orbits}, are consistent with those shown in Fig.~\ref{fig:leoII}. This highlights the relative insensitivity of Carina to tides, as well as the limited sensitivity to different halo initialization methods (Eddington versus halo eigenstate). For comparison, in the same figure we also show dashed curves corresponding to $r_{\rm t}=1$~kpc (green) and $r_{\rm t}=0.5$~kpc (dark yellow), which should be regarded as extreme possibilities for Leo~II (but not Carina). In such cases, the Leo~II data would become compatible with a ULDM halo with $m=\SI{5e-21}{\electronvolt}$, contrary to what was claimed in Ref.~\cite{Teodori:2025rul}. Nevertheless, we stress that, although tidal radii as small as $r_{\rm t}=1$~kpc or $0.5$~kpc are statistically allowed in the lower tail of our Monte Carlo orbit ensemble, they should be regarded as extreme possibilities for Leo~II. This is also because Leo~II does not show clear present-day evidence for strong tidal disturbance, either in its stellar morphology or internal kinematics~\cite{Coleman:2007xe, Koch:2007ye}. At the same time, the absence of obvious tidal features today does not by itself exclude stronger tidal effects in the past, since clear tidal signatures are observed only in a limited number of dwarf galaxies and can in general be difficult to detect.

For illustrative purposes, we show on the right panel of Fig.~\ref{fig:LeoII_5em21_tides} the energy eigenvalue cutoff for the three different tidal radii considered (see also Ref.~\cite{Eberhardt:2025lbx} for similar plots).

\section{Results: Leo I and Draco}\label{s:draco}
To further substantiate the tension for $m \lesssim 5\times10^{-21}$~eV, which has so far been robustly driven by Carina alone, we now analyze two additional dwarf galaxies, Leo~I and Draco. The tidal properties of these systems are also summarized in Tab.~\ref{tab:dwarf_inputs}. As far as the mass of the star component is concerned, Ref.~\cite{McConnachie2012} cites $M_* = 0.29\times 10^{6} M_\odot$ for Draco and $M_* = 5.5\times 10^{6} M_\odot$ for Leo I. We show in Fig.~\ref{fig:LeoI_2em21} an example of a run for $m=\SI{2e-21}{\electronvolt} $ targeting Leo I, with tidal radius $r_{\rm t} = \SI{1.5}{\kilo\parsec}$ and $r_{\rm t} = \SI{1}{\kilo\parsec}$,. As it is apparent, both the soliton-induced stellar kinematics peak and dynamical heating make $m=\SI{2e-21}{\electronvolt}$ incompatible with data. Although the mass of the stars is tailored for Leo I, results can equally be applied for Draco, which has a similar expected tidal radius and similar stellar kinematics, but lighter $M_*$ (which makes results conservative, as stars self-gravity would be less relevant). When targeting $m=\SI{5e-21}{\electronvolt}$ keeping fixed all the other parameters, dynamical heating does not drive the tension anymore, but the stellar kinematics inner peak remains prominent enough for both Leo I and Draco, see Fig.~\ref{fig:LeoI_5em21}. 

\section{Discussion} \label{s:disc}

Throughout this work, we have explicitly taken into account the effects of stellar self-gravity and tides on ULDM bounds based on dynamical heating. We show that both effects can modify the strength of dynamical heating, sometimes significantly. However, the main result of this work is that the mass range \(5\times 10^{-22} \lesssim m/\text{eV} \lesssim 5\times 10^{-21}\) remains excluded even when tides and stellar self-gravity are taken into account. In contrast to Ref.~\cite{Teodori:2025rul}, we have now also included Leo~I and Draco among our targeted dwarf galaxies. 

\begin{figure}
    \centering

\includegraphics[width=0.99\linewidth]{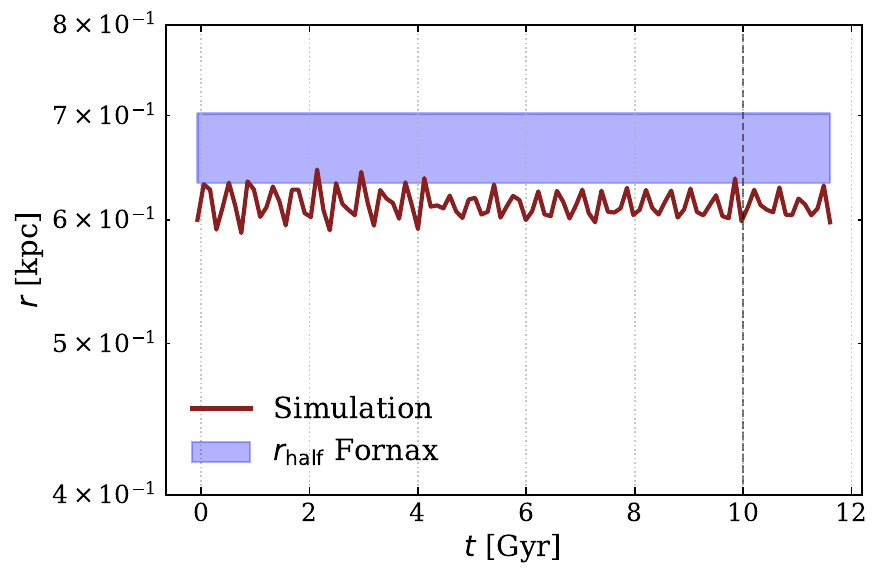}
    \caption{Fornax simulation, with tides, for $m=\SI{1e-22}{\electronvolt}$. The increase of the half-light radius is completely halted in this scenario, to be compared with Ref.~\cite{Yang:2025bae}. 
    }
    \label{fig:1em22_Fornax}
\end{figure}

A possible concern is our simplified implementation of tidal effects. We remark that, in a realistic setting, tidal stripping is a continuous process affecting the halo at all times, with time-dependent mass loss. Our approach is much simpler and more radical, since we assume from time zero that the halo has already been stripped down to the expected minimum tidal radius. A more reliable assessment of the impact of tidal stripping will require more refined and realistic simulations. Nevertheless, when comparing our results for \(m=\SI{1e-22}{\electronvolt}\), shown in Fig.~\ref{fig:1em22_Fornax}, with the strong-tide scenario of Ref.~\cite{Yang:2025bae}, we find that in our case the stars are not affected by heating at all, while they are for Ref.~\cite{Yang:2025bae}. This suggests that our method is conservative. We are, however, aware that a proper assessment will require a better characterization of tidal effects, which should be implemented in future work. For example, the tidal field of the host halo can also affect the stellar distribution, as well as the gas distribution~\cite{Williamson_2016}, and not only the dark matter one. This could in some cases lead to a further increase in the half-light radius~\cite{Errani_2015}, making our bounds even more conservative, although the evolution of the half-light radius may depend sensitively on the orbital phase and the dark matter profile~\cite{Sanders_2018}. Tides can also decrease the stellar dispersion velocity (if they reduce the enclosed mass), and their precise impact on the stellar distribution therefore deserves a dedicated study. It is also worth remarking that our simulations are agnostic about the actual formation history of the dwarf galaxy halo 
and that for the masses of interest the core should not be affected by tunneling effects due to tides~\cite{Hertzberg:2022vhk}. Along these lines, we emphasize that the initial time of our simulations should not  be identified with the formation time of the dwarf galaxy. The systems  considered here may have spent a non-negligible period evolving outside  the Milky Way virial radius before first infall. During this pre-infall 
phase, the stellar component could have been shaped by internal  processes before Milky Way tides became important. We do not attempt to model this full isolated galaxy-formation 
history. Instead, our goal is to isolate the effect of ULDM-induced  dynamical heating, together with two mechanisms that can reduce it, stellar self-gravity and tidal stripping. From this perspective, our treatment of tides is conservative. In the  runs where tides are included, the ULDM halo is initialized as already 
truncated at the adopted tidal radius, which we take from the smallest  tidal radius inferred along the orbit. A real dwarf evolving in  isolation before Milky Way infall would instead have an unstripped, more extended halo. In the eigenmode picture, such a halo contains a 
larger population of weakly bound, spatially extended modes, whose  interference contributes to the time-dependent density fluctuations  responsible for ULDM heating. Therefore, including an explicit isolated 
pre-infall phase would not introduce tidal suppression during that  period and would generally increase the accumulated ULDM heating within 
our framework.

Throughout this work, we have used the expected stellar mass of each system to inform our simulations, and we have used the estimates in Tab.~\ref{tab:dwarf_inputs} to set the initial \(r_{\rm t}\). Our results could be affected if one or both of these estimates are incorrect, in particular if the \emph{initial} stellar mass is much larger than expected and/or if our estimate of the tidal radius is biased high. As possible examples, in Fig.~\ref{fig:leoII} we have shown that dynamical-heating constraints are significantly weakened for Leo~II/Carina if the actual stellar mass is five times larger than that inferred from mass-to-light ratio studies, and in Fig.~\ref{fig:LeoII_5em21_tides} we see the same weakening effect if the appropriate \(r_{\rm t}\) for Carina were smaller than \(2\)~kpc. We however remark, as pointed out in Sec.~\ref{s:res:ForCarLeo}, that such large variations of stellar mass and tidal radius determinations are statistically very unlikely, at least for Carina. By contrast, the results based on the kinematic dispersion peaks for Fornax, Leo I and Draco are more robust and largely unaffected by stellar mass determination and tidal stripping caveats. 

Our limits are complementary to, but weaker than, those derived from the suppression of the subhalo mass function in recent cosmological zoom-in simulations~\cite{Nadler:2025fcv} (see also Ref.~\cite{Liu:2025vhk, May:2022gus}). However, these limits are obtained under the assumption that the only effects of fuzzy dark matter on structure formation are encoded in a modified transfer function and power spectrum. This is likely an idealization that does not capture all the relevant physics (see for example the discussion in Section 5 of Ref.~\cite{Hui:2021tkt}). The ultimate, albeit very ambitious, goal of the field should be to carry out cosmological simulations that capture the full ULDM dynamics, from high redshift to the present day, including both the impact on the subhalo mass function and on stellar dynamics in dwarf galaxies.

Finally, ultra-faint dwarf galaxies such as Segue~I provide access to even larger ULDM masses, in the range $\sim 10^{-20}$--$10^{-19}~\mathrm{eV}$. Refs.~\cite{Dalal:2022rmp,May:2025ppj} claim to exclude ULDM masses $m \lesssim 3 \times 10^{-19}~\mathrm{eV}$. However, in upcoming work~\cite{TeodoriCaputoUltra2_inprep} we will show that the more likely excluded range---if Segue~I is indeed dark-matter dominated, which is not yet clear~\cite{lujan2025darkmatterdominatedgalaxysegue}---is instead $10^{-20} \lesssim m \lesssim 10^{-19}~\mathrm{eV}$. This highlights that one should study both classical dwarfs and ultra-faint dwarfs in order to probe different ULDM mass ranges (see also Ref.~\cite{Yang:2026wdk} for a study of Segue I in the presence of $m=\SI{1e-22}{\electronvolt}$ ULDM).

\section{Conclusions} \label{s:conc}
The presence of inner cored profiles, and dynamical heating, are defining phenomenological features of ULDM, both of which can be put to test with galaxy data. In particular, dwarf galaxies are the most susceptible to $m \gtrsim 10^{-22}$ eV ULDM. In Ref.~\cite{Teodori:2025rul}, we concluded that $ 5\times 10^{-22} \lesssim m/\text{eV} \lesssim 5\times 10^{-21} $ is in tension with stellar kinematics and stellar distribution data of Fornax, Carina and Leo II dwarf galaxies. Here we expanded on the results of Ref.~\cite{Teodori:2025rul}, by explicitly looking at the influence of star self gravity and tidal stripping on dynamical heating in dwarf-galaxy systems like the aforementioned Fornax, Leo II, Carina, with the further addition of Leo I, Draco. We confirm the tension of ULDM with dwarf galaxy stellar kinematics and distribution data for $ 5\times 10^{-22} \lesssim m/\text{eV} \lesssim 5\times 10^{-21} $, compatibly with the conclusions of Ref.~\cite{Teodori:2025rul}. Nevertheless, we here showed that both star self-gravity and tidal stripping can affect the dynamical heating of ULDM in significant ways, as also pointed out in Ref.~\cite{Eberhardt:2025lbx}. All future works discussing ULDM dynamical heating on dwarf galaxies will have to explicitly take both into account. As a next step, it will be important to study tidal interactions more precisely by evolving the ULDM halo within the Milky Way gravitational potential. However, such an approach would still remain subject to important modeling and numerical uncertainties, including the treatment of mass loss, numerical resolution, the assumed Milky Way potential, and the practical difficulty of performing broad scans over ULDM masses and dwarf initial conditions. In this sense, the simplified setup adopted here and in Ref.~\cite{Eberhardt:2025lbx} should not only be viewed as an approximation, but also as a more controlled and computationally tractable framework against which future, more elaborate simulations can be compared.

\vspace{0.5cm}

\textit{Acknowledgments.} We are grateful to K. Blum for many useful conversations and comments on the draft, and to E. Hardy and M. Gorghetto for the early development of a version of the ULDM solver used in this work. We also thank A. Eberhardt, Y.M. Yang, Z.C. Zhang, X.J. Bi, and P.F. Yin for comments on the draft of this work. The authors acknowledge the support by the European Research Area (ERA) via the UNDARK project (project number 101159929). AC is also supported by an ERC STG grant (``AstroDarkLS'', grant No. 101117510). AC also acknowledges the Weizmann Institute of Science for hospitality at different stages of this project and the support from the Benoziyo Endowment Fund for the Advancement of Science. LT acknowledges the MICINN through the grant ``DarkMaps'' PID2022-142142NB-I00.

\appendix

\section{Code}
\label{s:code}
We employ an updated version of the Ref.~\cite{Teodori:2025rul} pseudo-spectral 3D Schr\"odinger-Poisson Equation (SPE) solver with stars. The SPE solver closely resembles the one described in~\cite{Levkov:2018kau}, and different versions of the same SPE solver implementation used in this work, which do not include star dynamics, were used in~\cite{Blum:2024igb,Gorghetto:2022sue,Budker:2023sex,Gorghetto:2024vnp}. In the following, we review the basics of its implementations.

The SPE for the field $ \psi $ with mass $ m $ read
\begin{align} \label{eq:SPEfield}
	&\iu\pdv{\psi}{t} = -\frac{1}{2m} \laplacian{\psi} + m(\Phi + \Phi_{\rm s})\psi \ , \\
& \laplacian \Phi = 4\pi G(|\psi|^2 - \langle|\tilde\psi|^2|\rangle) \ ,	
\end{align}
where $\Phi$ is the gravitational potential sourced by the ULDM field $\psi$, whereas $\Phi_{\rm s}$ is an external potential, represented by the stars in our case.
In this convention, $ \psi $ has the dimensions of a mass squared.

The ULDM field is evolved via the unitary operator
\begin{equation}
\psi(t+ \dd t) = \prod_{\alpha} \e^{-\iu d_\alpha m\dd{t} (\Phi_\alpha + \Phi_{\rm s})} \e^{-\iu c_\alpha\dd{t} \frac{(-\iu \grad_{x})^2}{2m} }\psi(t) \ ;
\end{equation}
the equation is meant to be read from right to left, i.e. before one applies the kinetic operator
\begin{equation}\label{eq:kin_op}
\e^{-\iu c_\alpha\dd{t} \frac{(-\iu \grad_{x})^2}{2m} }\psi(t) =: \psi^{(\alpha)}(t +\dd{t} ) \ ,
\end{equation}
and then the potential operator, where
\begin{equation}\label{eq:Poiss_num}
\laplacian_{x}{\Phi_\alpha} = |\psi^{(\alpha)}|^2 \ .
\end{equation}
The Poisson equation here is sourced by the ULDM field alone. In practice, the kinetic operator is most conveniently applied in Fourier space, whereas the nonlinear potential term is evaluated in position space~\cite{Levkov:2018kau}; the evolution is therefore implemented by alternating between the two representations using Fast Fourier Transforms (FFT) with the FFTW library~\cite{10.1145/301631.301661}. The constants $ c_\alpha $, $ d_\alpha $ can be found in~\cite{Levkov:2018kau}.

Star dynamics is computed from the gravitational potential $\Phi + \Phi_{\rm s}$. Star coordinates are evolved via the leapfrog integrator~\cite{2008gady.book.....B} ($\Phi_{\rm tot} := \Phi + \Phi_{\rm s}$)
\begin{align}
\begin{aligned}
    &\vec{r}_i(t_{j+1/2}) = \vec{r}_i(t_j) + \frac{1}{2} \dd{t} \vec{v}_i(t_j) \ , \\ 
    &\vec{v}_i(t_{j+1}) = \vec{v}_i(t_j) - \dd{t} \grad\Phi_{\rm tot}(\vec{r}_i(t_{j+1/2})) \ ; \\ 
    &\vec{r}_i(t_{j+1}) = \vec{r}_i(t_{j+1/2}) + \frac{1}{2} \dd{t} \vec{v}_i(t_{j+1/2}) \ .
\end{aligned}
\end{align}
$\Phi_{\rm s}$ is computed solving for the stellar Poisson equation in the grid via FFT, where the mass density of stars is assigned to the grid points via the Cloud In Cell (CIC) algorithm.

The gradient of $\Phi_{\rm tot}$ is computed via the 5 point midpoint numerical scheme, and its value at the star coordinates (which generically are subgrid) is recovered via multi-linear interpolation. Notice that multi-linear interpolation is complementary to the CIC mass assignment. With this method, single stars feel the gravitational potential of themselves. To correct for this spurious effect, at the beginning of the simulation, we compute the self force caused by a single star located in all the different points of an equally spaced $4^3$ subgrid. For a stars in a subgrid position, we interpolate the self force from this precomputed table and subtract it from the force obtained via the potential gradient.

Coherently with the leapfrog integrator, we evolve ULDM for $\dd t/2$, perform the drift step of stars, then perform the kick with the $t+\dd t/2$ potential for both drifted stars and $t+ \dd t$ evolved ULDM.

%

The total energy of the system reads 
\begin{align}
\begin{aligned}
	E_{\rm tot} &= \int \dd[3]{x} \qty( \frac{|\grad{\psi}|^2}{2m^2} + \frac{1}{2} |\psi|^2 \Phi_{\rm tot} ) \\
    &+ \sum_{j\in \text{stars}} \qty(\frac{m_j}{2} v^2_j +m_j\Phi_{\rm tot}(\vec{r}_j))\ .
\end{aligned}
\end{align}

We use adaptive time steps, to ensure conservation of energy $ \Delta E_{\rm tot}/E_{\rm tot} \lesssim 10^{-5} $ between time-steps, and an overall conservation of energy $ |E_{\rm tot}^{\rm final} - E_{\rm tot}^{\rm initial}|/|E_{\rm tot}^{\rm final} + E_{\rm tot}^{\rm initial}| \lesssim 10^{-3} $.

Further details on the code implementation and validation can be found in Ref.~\cite{Teodori:2025rul}.

\section{Eigenmode solver} \label{s:eigen}
We here show some details of the eigenmode solver of Ref.~\cite{Yavetz:2021pbc}, used to inform the initial conditions of our simulations where tidal stripping is taken into account as a cutoff on the energy of such modes.

Given a spherically symmetric gravitational potential $\Phi(r)$ of the target profile $\rho(r)$ we want our ULDM field $\psi$ to reproduce, we can consider a spherical harmonic decomposition to express the $\psi$ eigenfunctions
\begin{equation}
\psi_{nlm}(r,\theta,\phi) = \frac{u_{nl}(r)}{r} Y_{lm}(\theta, \phi) \ ,
\end{equation}
where the radial functions $u_{nl}(r)$, for each $n,l$, solve the eigenvalue equation 
\begin{align}
\begin{aligned}
\frac{-\hbar^2}{2m} \dv{^2 u_{nl}}{r^2} + \qty(\frac{\hbar^2}{2m} \frac{l(l+1)}{r^2} + m \Phi(r)  ) u_{nl} = E_{nl} u_{nl} \ ,
\end{aligned}
\end{align}
where $n$ denotes the number of nodes of $u_{nl}$. $Y_{lm}$ are spherical harmonics.

\begin{figure}
    \centering
\includegraphics[width=1.15\linewidth]{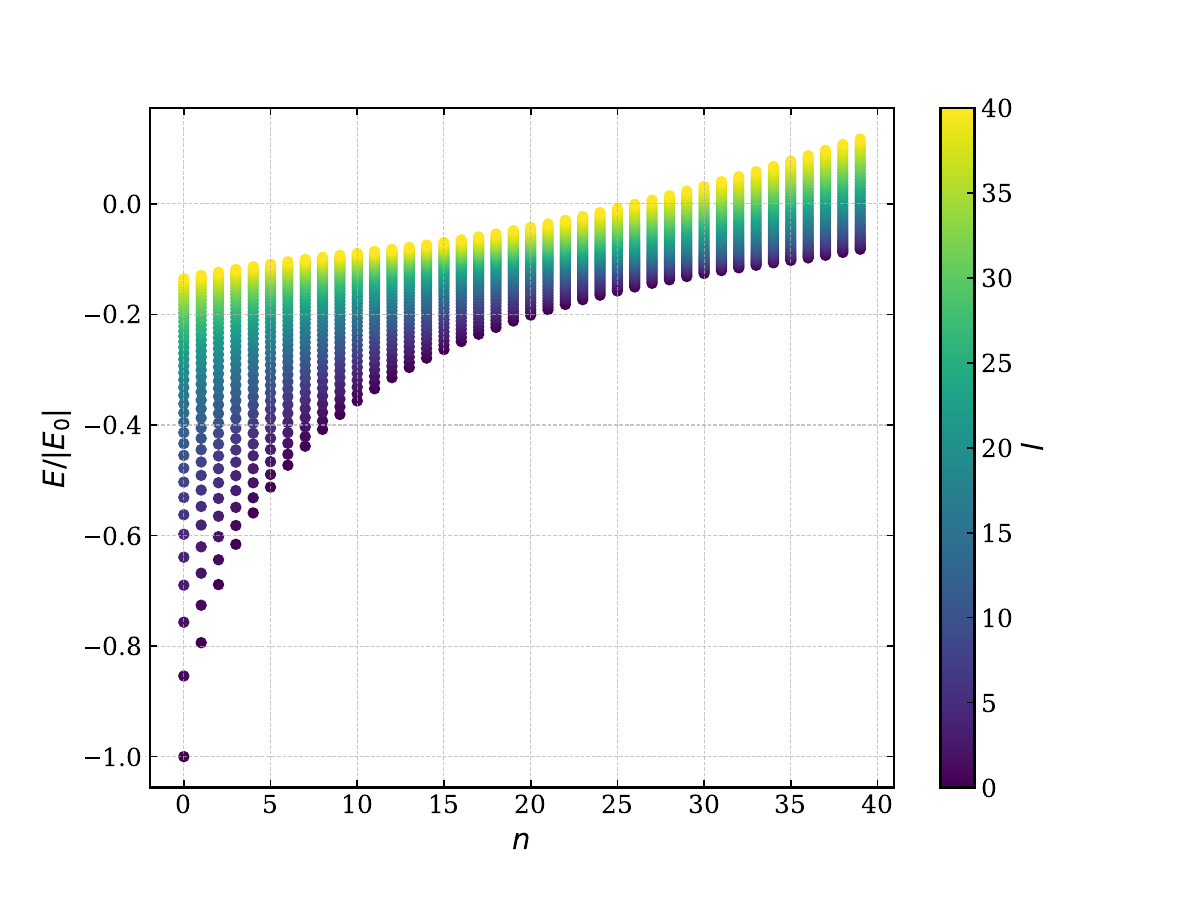}
    \caption{Energy of eigenstates of a given pair ${n,l}$, normalised to the energy of the ground state, $E_0$. Notice the weaker dependence on $l$.}
    \label{fig:Energy}
\end{figure}

Assuming isotropy (no preferred direction), we take the statistical properties of the modes to be independent of $m$ at fixed $(n,l)$. We then match the target \emph{spherically averaged} density profile by choosing non-negative weights $C_{nl}$ such that
\begin{equation}
\rho(r)\simeq \sum_{n,l} C_{nl}\,\frac{|u_{nl}(r)|^2}{r^2}\,;
\end{equation}
a convenient construction that guarantees this matching (after averaging over random phases) is
\begin{equation} \label{eq:sum_eigen}
\psi(\mathbf{x},t)=
\sum_{n=0}^{n_{\rm max}}\sum_{l=0}^{l_{\rm max}}\sum_{m=-l}^{l}
\sqrt{\frac{4\pi\,C_{nl}}{2l+1}}\;
\psi_{nlm}(\mathbf{x})\,e^{i\varphi_{nlm}}\,e^{-iE_{nl}t/\hbar}\ ,
\end{equation}
where the $\varphi_{nlm}$ are independent random phases.
Indeed, using $\sum_{m=-l}^{l}|Y_{lm}(\theta,\phi)|^2=(2l+1)/(4\pi)$, one finds
\begin{equation}
\left\langle |\psi(\mathbf{x},t)|^2 \right\rangle_{\varphi}
=\sum_{n,l} C_{nl}\,\frac{|u_{nl}(r)|^2}{r^2}\ ,
\end{equation}
while the off-diagonal ($n\neq n'$ and/or $l\neq l'$) interference terms, which oscillate at frequencies set by energy differences, produce the time-dependent density fluctuations responsible for dynamical heating. A tidally stripped halo is then represented as in Eq.~\eqref{eq:sum_eigen}, with a cutoff on the pair $(n_{\rm max}, l_{\rm max})$ such that the resulting density profile does not have support for radii $r \gtrsim r_{\rm t}$, where $r_{\rm t}$ is the tidal radius. In Fig.~\ref{fig:Energy} we also show the corresponding energy for a given eigenstate, normalised to the ground state energy.

\section{Dwarf-galaxy orbital histories} \label{s:orbits}
We here describe how we obtain the tidal histories of the dwarf galaxies we focus on and the values in Tab.~\ref{tab:dwarf_inputs}.

We model the Milky Way gravitational field with two axisymmetric potentials implemented in the \texttt{galpy} dynamics library \citep{Bovy2015galpy}.
As a fiducial model we adopt the \texttt{McMillan17} Milky Way potential \citep{McMillan2017}, which is calibrated
to a broad compilation of Milky Way constraints. We also checked our results using the \texttt{MWPotential2014} potential, a three-component
bulge--disk--halo potential recommended in \texttt{galpy}~\cite{Bovy2013ApJ}, but we then decided to adopt the \texttt{McMillan17} potential also because it is a conservative choice for the scopes of this work (tides are more sizable within this potential). For each dwarf galaxy we specify its present-day sky position $(\alpha,\delta)$,
heliocentric distance $D$, line-of-sight velocity $v_{\rm los}$, and proper motions
$(\mu_{\alpha*},\mu_{\delta})$~\cite{Battaglia_2022}. We treat $(D,v_{\rm los},\mu_{\alpha*},\mu_{\delta})$
as Gaussian-distributed about their measured central values with quoted $1\sigma$
uncertainties, while $(\alpha,\delta)$ are held fixed. We generate $N_{\rm MC}$ Monte Carlo realisations of the present-day observables
to propagate measurement uncertainties into orbital parameter posteriors.

\begin{figure}
    \centering
    \includegraphics[width=0.99\linewidth]{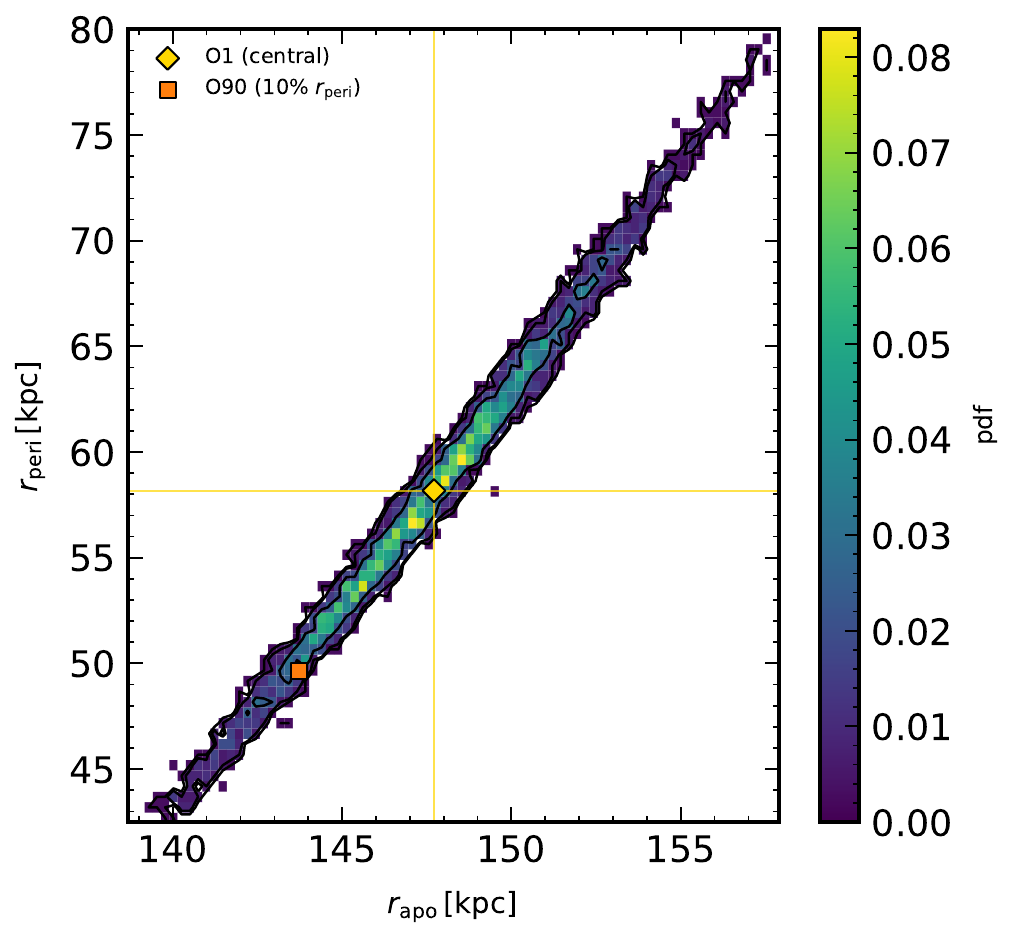}
    \caption{Monte Carlo posterior for Fornax’s orbital apocenter and pericenter in the McMillan Milky Way potential~\cite{McMillan2017}. We draw $N=3000$ realizations of $(D_\odot, v_{\rm los}, \mu_{\alpha*}, \mu_\delta)$ from their quoted uncertainties, integrate each orbit backward for $10\,\mathrm{Gyr}$, and compute $(r_{\rm apo}, r_{\rm peri})$ from the extrema of the Galactocentric radius $r(t)$. The color map shows the estimated probability density in the $(r_{\rm apo}, r_{\rm peri})$ plane; we highlight the fiducial (gold diamond) orbit and the low-pericenter quantile orbit (orange square), more plunging than $90 \%$ of the orbits.
}
    \label{fig:PDFPeriFornax}
\end{figure}

We then integrate each orbit backward in time for $T=10~{\rm Gyr}$ using the
\texttt{galpy.orbit.Orbit} class and use a uniform time grid $t\in[0,-T]$ with $N_t$ steps and the symplectic integrator
\texttt{symplec4\_c}. From the integrated phase-space trajectory we compute the Galactocentric distance
$r(t)$ and derived pericentre and
apocentre radii by the discrete extrema on the integration grid,
\begin{equation}
r_{\rm peri} \simeq \min_{t} r(t) \ ,
\qquad
r_{\rm apo} \simeq \max_{t} r(t) \ .
\label{eq:rperi_rapo_discrete}
\end{equation}
We have verified that increasing $N_t$ does not materially change the resulting
$r_{\rm peri}$ distributions for the systems studied. As an example, in Fig.~\ref{fig:PDFPeriFornax} we show the distribution of Fornax apocentric and pericentric radii from our orbit integrator in the McMillan (2017) Milky Way potential, which is in good agreement with Fig.~3 of Ref.~\cite{Di_Cintio_2024}. 

\subsection{Tidal radius and tidal-to-stellar size ratio}
\label{sec:tidal_radius}
To quantify tidal susceptibility along the orbit, we compute an instantaneous tidal radius from the gravitational tidal tensor of the host potential,
\begin{equation}
  r_{\rm t}^{3} = \frac{G\,m(<r_{\rm t})}{\Lambda_{\rm eff}},
  \qquad
  \Lambda_{\rm eff} \equiv \lambda_{\max} + \Omega_{\rm inst}^2,
  \label{eq:rt_tensor}
\end{equation}
where $m(<r)$ is the satellite enclosed mass profile~\cite{Walker:2009zp}, $\lambda_{\max}$ is the largest eigenvalue of the tidal tensor $T_{ij}=-\partial^2\Phi_{\rm host}/\partial x_i\partial x_j$ evaluated at the satellite's instantaneous position, and
$\Omega_{\rm inst}$ is the instantaneous angular velocity of the satellite about the Galactic centre. The eigenvalues are computed with \texttt{galpy}'s \texttt{ttensor} routine applied to the McMillan potential~\cite{McMillan2017}. Compared with the classical Roche estimate $r_{\rm t}=R[m/(3M_{\rm host})]^{1/3}$, Eq.~\eqref{eq:rt_tensor} properly accounts for the local shape of the host potential — including the contribution of the local density to the tidal field — and uses the satellite's true angular velocity rather than the circular-orbit value. Equation~\eqref{eq:rt_tensor} is solved iteratively for $r_{\rm t}$ at each time step, assuming the satellite follows an NFW profile with parameters $(M_{200},c)$, and is capped at $r_{200}$ when the tidal field is too
weak to strip any material. Given the observational uncertainties on the present-day phase-space coordinates and our approximate treatment of tidal effects on ULDM, this level of modeling is sufficient for our purpose of assessing the typical tidal susceptibility and its uncertainty.
For the satellite mass profile we adopt an NFW halo, parameterised by $(M_{200},c)$.
Writing $r_{\rm s}=r_{200}/c$ and
\begin{equation}
f(x) = \ln(1+x) - \frac{x}{1+x} \ ,
\label{eq:nfw_fx}
\end{equation}
the enclosed mass is
\begin{equation}
m(<r) = M_{200}\,\frac{f(r/r_{\rm s})}{f(c)} \ .
\label{eq:nfw_menc}
\end{equation}

To get an idea of the relevant scales, one can use the standard Roche estimate as a simple back-of-the-envelope calculation,
\begin{align}
\begin{aligned}
r_{\rm t}(r_{\rm peri})
&\sim r_{\rm peri} \left[\frac{m(<r_{\rm t})}{3\,M_{\rm host}(<r_{\rm peri})}\right]^{1/3} \\
&\simeq5.6~\mathrm{kpc}\, \left(\frac{r_{\rm peri}}{60~\mathrm{kpc}}\right)
  \left[\frac{m(<r_{\rm t})}{10^{9}~M_\odot}\right]^{1/3} \\
  &\quad\times\left[\frac{M_{\rm host}(<r_{\rm peri})}{4\times 10^{11}~M_\odot}\right]^{-1/3},
\end{aligned}
\label{eq:rt_roche_peri}
\end{align}
and therefore
\begin{align}
\begin{aligned}
\frac{r_{\rm t}(r_{\rm peri})}{r_{1/2}} \sim& \, 8 \, \left(\frac{0.7\, \text{kpc}}{r_{1/2}}\right)\left(\frac{r_{\rm
peri}}{60~\mathrm{kpc}}\right) \left[\frac{m(<r_{\rm t})}{10^{9}~M_\odot}\right]^{1/3} \\
&\times\left[\frac{M_{\rm host}(<r_{\rm peri})}{4\times 10^{11}~M_\odot}\right]^{-1/3}.
\end{aligned}
\end{align}
We stress that this is only a rough estimate; our numerical results are based on the full gravitational tidal tensor of the host potential discussed above, which can differ from the simple Roche formula by up to $\sim\!25\%$.

\begin{figure*}[t]
    \centering
    \begin{subfigure}{0.49\textwidth}
        \centering
        \includegraphics[width=\linewidth]{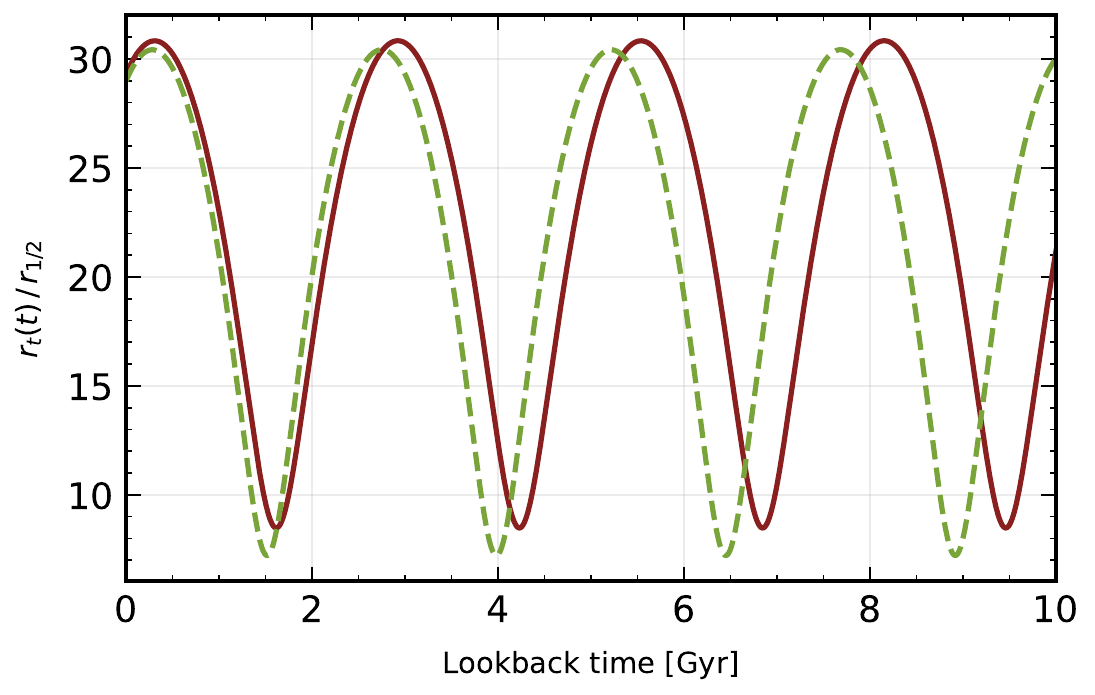}
        \caption{Fornax.}
    \end{subfigure}
    \hfill
    \begin{subfigure}{0.49\textwidth}
        \centering
        \includegraphics[width=\linewidth]{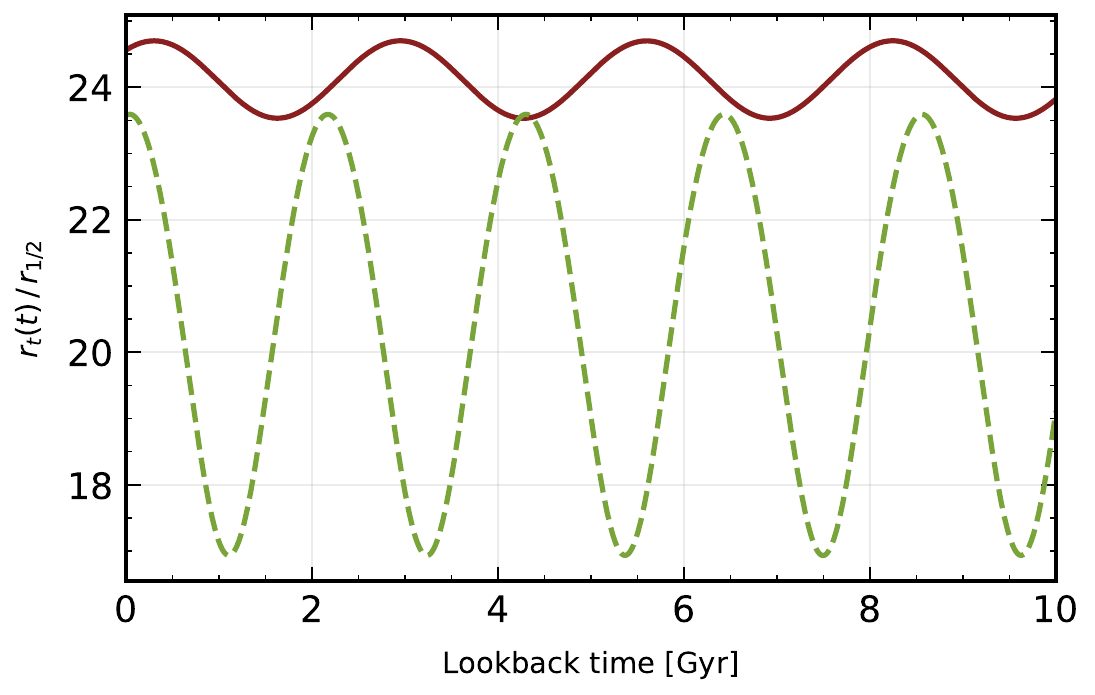}
        \caption{Carina.}
    \end{subfigure}
    \caption{Monte Carlo–based tidal evolution of Fornax (left) and Carina (right) in the
  McMillan Milky Way potential~\cite{McMillan2017}. The solid curve shows the
  tidal-to-half-light ratio $r_{\rm t}(t)/r_{1/2}$ along the fiducial orbit (O1;
  central values of $D_\odot$, $v_{\rm los}$, $\mu_{\alpha*}$, $\mu_\delta$).
  The dashed curve shows the same quantity for a low-pericenter realization
  selected from the Monte Carlo ensemble by requiring its pericenter to lie at
  the 16th percentile of the $r_{\rm peri}$ distribution. The tidal radius is
  obtained from the eigenvalues of the tidal tensor $T_{ij} =
  -\partial^2\Phi_{\rm host}/\partial x_i\partial x_j$ evaluated along the
  orbit, balancing the tidal stretching against the satellite's self-gravity;
  the satellite is modeled as an NFW halo with parameters $(M_{200},c)$. Time is
   shown as lookback time. We notice that the larger separation between fiducial and low-pericenter curves for Carina reflects the near-circularity of its fiducial orbit: small proper-motion
  perturbations redistribute angular momentum efficiently, producing large fractional changes in pericenter while leaving the apocenter nearly unchanged. This contrasts with Fornax, which already has moderate eccentricity and thus the same fractional proper motion error causes a smaller relative change in orbital shape.}
    \label{fig:RtRh_Fornax_Carina}
\end{figure*}

In Fig.~\ref{fig:RtRh_Fornax_Carina} we show the tidal-to-half-light ratio $r_{\rm t}(t)/r_{1/2}$ for Fornax (left) and Carina (right). The solid red curve is computed along the fiducial orbit (central values of $D_\odot$, $v_{\rm los}$, $\mu_{\alpha*}$, $\mu_\delta$). The dashed curve shows the same quantity for a more plunging orbit selected from the Monte Carlo ensemble by requiring its pericenter to lie at the 16th percentile of the $r_{\rm peri}$ distribution (i.e.\ a ``$+1\sigma$'' low-pericenter realization). For Fornax, we find $\min_t\!\left[r_{\rm t}(t)/r_{1/2}\right]\simeq 8$--$9$, implying that the stellar component is well within the instantaneous tidal radius, while the outer dark-matter halo (which extends to radii $\gg r_{1/2}$) can still be susceptible to tidal stripping. Our inferred tidal radius is larger by a factor of $\sim 2$ than that reported in Ref.~\cite{Yang:2025bae}. A likely origin of this difference is the assumed halo mass: Ref.~\cite{Yang:2025bae} adopt $M_{200}=4\times 10^8\,M_\odot$ for Fornax, whereas we use a pre-infall mass of order $M_{200}\sim (1-2)\times 10^{10}\,M_\odot$, as typically inferred from abundance-matching--based estimates for Fornax-like systems~\cite{Read:2018fxs}. For Carina, the situation seems better although the pericenter value is quite sensitive to orbital parameters. This is due to the fact that Carina's orbit is near-circular and small proper-motion perturbations
redistribute angular momentum efficiently, producing large fractional changes in pericenter (while leaving the apocenter nearly unchanged).

In Tab.~\ref{tab:dwarf_inputs}, we summarize the orbital parameters and tidal features of a selected sample of dwarfs. We report the median value of $\min_{t}\!\left[\frac{r_{\rm t}(t)}{r_{1/2}}\right]$ together with the 2.5th--97.5th percentile intervals, obtained by propagating the observational errors on distance, line-of-sight velocity, and proper motions
via Monte Carlo sampling. For each realization we draw $(D_\odot, v_{\rm los}, \mu_{\alpha*}, \mu_\delta)$ from independent Gaussian distributions centered on the measured values with widths equal to the quoted uncertainties, integrate the orbit in the McMillan potential, and evaluate $\min_{t}\!\left[\frac{r_{\rm t}(t)}{r_{1/2}}\right]$ along the resulting trajectory. The tidal radius at each point is computed from the eigenvalues of the tidal tensor $T_{ij}=-\partial^2\Phi_{\rm host}/\partial x_i\,\partial x_j$, balancing the dominant stretching eigenvalue against the satellite's self-gravity modeled as an NFW halo.

\section{Simulations details} \label{s:details}
In Tab.~\ref{tab:details}, we show parameters and details of our simulation suite. The halo parameters, $r_{\rm s}$ and $\rho_{\rm s}$ or $r_{\rm c}^{\rm ini}$ for simulations without and with tides implemented respectively, are chosen to give a stellar kinematics curve which gives a match as close as possible with data (for settings where the soliton-induced inner peak would make it impossible to match the full stellar kinematics curve, we tried to match the tail of such curve). Varying these parameters such that the stellar kinematics curve is not affected, leaves our results unchanged, as also showed in Ref.~\cite{Teodori:2025rul}. The choice of $r_{\rm t}$ follows Table~\ref{tab:dwarf_inputs}, whereas the total mass of the stars follows what is expected from the target system. The choice of the length of the box comes from a compromise between the expected virial radius of the system (of order $15 - 20$ kpc for the dwarf galaxies considered here) and the resolution needed to resolve the stars and ULDM dynamics. We tested that increasing the length of the box while maintaining the resolution does not change results.

\begin{figure*}
    \centering
\includegraphics[width=0.49\textwidth]{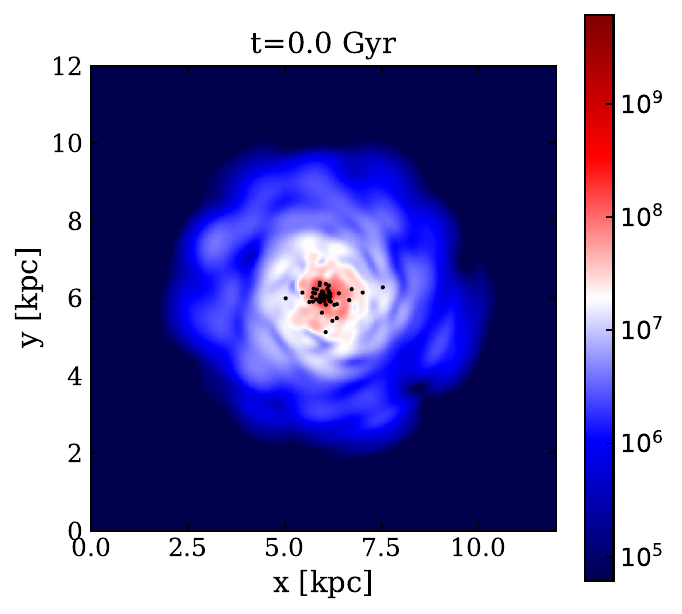}
\includegraphics[width=0.49\textwidth]{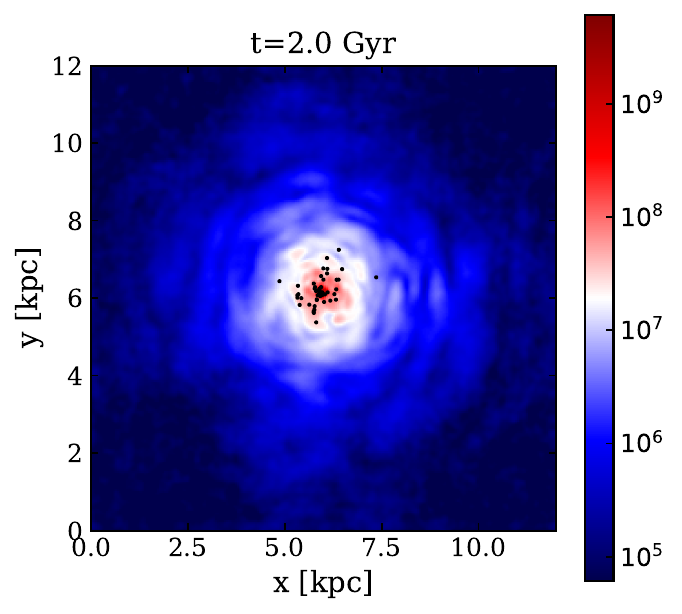}
 \includegraphics[width=0.49\textwidth]{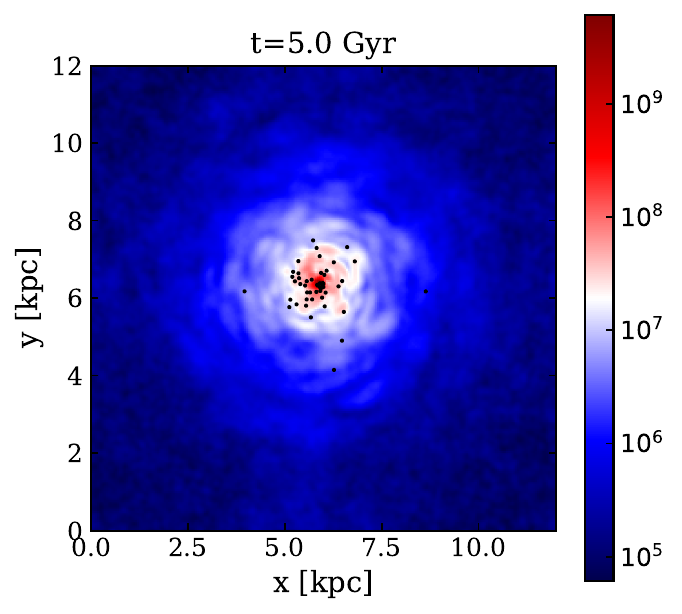}
\includegraphics[width=0.49\textwidth]{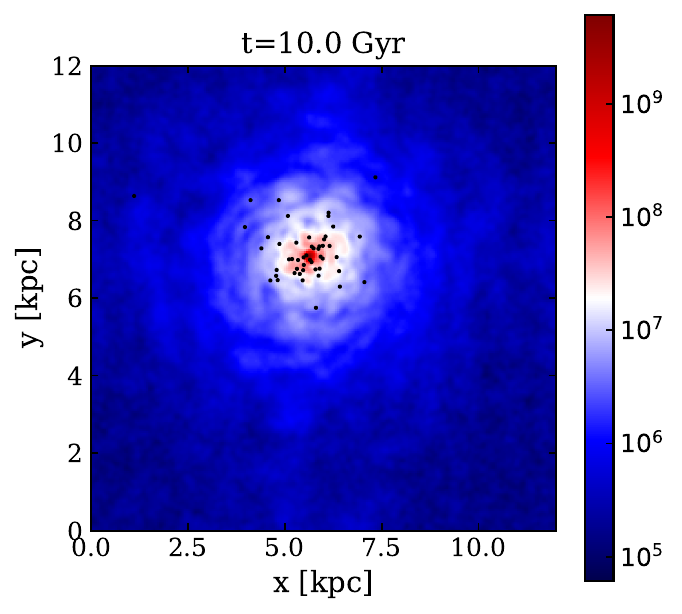}
\caption{Column density snapshots (in $M_\odot /$kpc$^2$) of a simulation with $m=\SI{e-21}{\electronvolt}$ in a box of length $L = 12$ kpc, whose main results are shown in Fig.~\ref{fig:Fornax_1em21}. Black dots show 100 randomly selected stars (out of 10000 stars in this simulation).}
\label{fig:snaps}
\end{figure*}

\begin{figure*}
    \centering
    \includegraphics[width=0.49\linewidth]{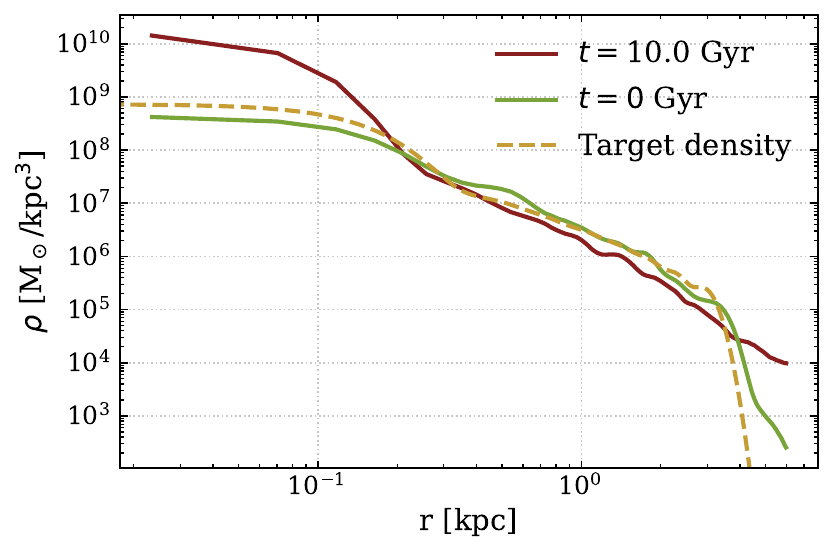}
    \includegraphics[width=0.49\linewidth]{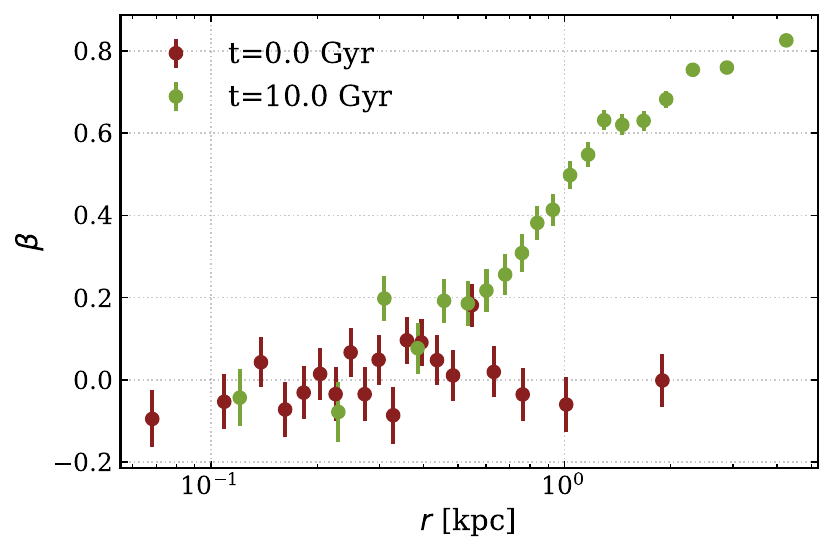}
    \includegraphics[width=0.49\linewidth]{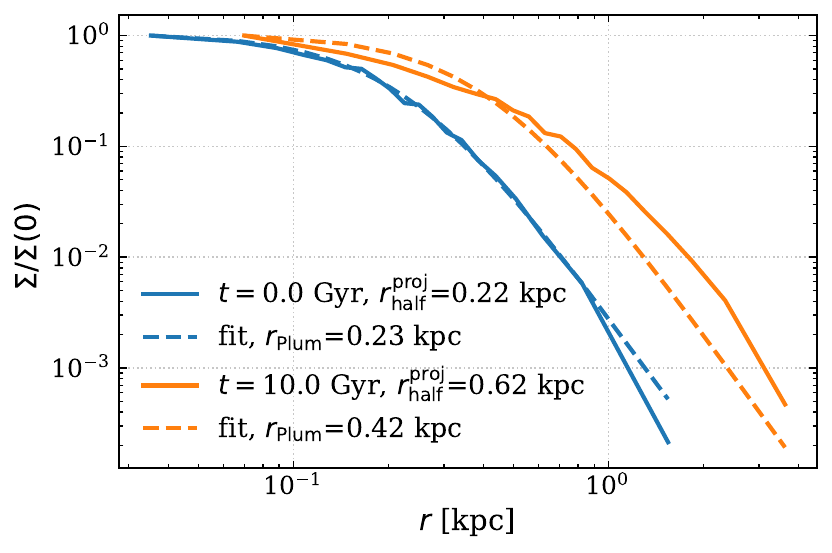}
    \caption{Some more plots for the simulation in Fig.~\ref{fig:Fornax_1em21}, taken as example. {\bf Top left}: radially averaged density profile for $t=0$ Gyr and $t=10$ Gyr. In dashed yellow, we show the initial target profile, found with the eigenvalue procedure outlined in the main text. It differs from the $t=0$ Gyr one, since we let evolve the ULDM field for $\sim 1$ Gyr before inserting stars to wait out for possible initial condition transients.  
    {\bf Top right}: Radial behavior of the anisotropy parameter $\beta$. 
    {\bf Bottom}: column density $\Sigma$ profile, compared with a Plummer fit, dashed line.}
    \label{fig:plots}
\end{figure*}

For illustrative purposes, in Fig.~\ref{fig:snaps} we show column density snapshots of the simulation described in Fig.~\ref{fig:Fornax_1em21}, here taken just as an example. In Fig.~\ref{fig:plots} we show instead dedicated plots of the same simulation.

\begin{table*}
\centering
\begin{tabular}{ccccccccccccc}
\toprule
$\displaystyle \frac{m}{\text{eV}}$  
& Target 
& $\displaystyle\frac{L}{{\rm kpc}}$ 
& $\displaystyle\frac{r_{\rm s}}{{\rm kpc}}$ 
&$\displaystyle\frac{\rho_{\rm s}}{10^7 M_\odot/{\rm kpc^3}} $ 
& $\displaystyle\frac{r^{\rm ini}_{\rm c}}{10\mathrm{pc}} $
& $\displaystyle\frac{r^{\rm ini}_{\rm Plum}}{{\rm kpc}}$ & $\displaystyle\frac{r_{\rm half}^{10\, {\rm Gyr}}}{{\rm kpc}}$ & $\displaystyle\frac{\sigma^{10\, {\rm Gyr}}_{\rm los}}{{\rm km/s}}$ & $\displaystyle \frac{r^{10 \,{\rm Gyr}}_{\rm c}}{10 \,\mathrm{pc}}$  & $\displaystyle \frac{r_{\rm t}}{\rm kpc} $ & $\displaystyle \frac{M_*}{10^6 M_\odot}$ & Figure \\
\midrule
$ 1 \times 10^{-22}$ & Fornax & $15$ &  $1.5$ & $N/A$ & 83.64 & $0.3$ & $0.21^{+0.03}_{-0.00}$ & $17.9^{+2.2}_{-0.4}$ & $58.0^{+13.3}_{-13.3}$ & $3.00$ & $113.68 $ & \\
$ 1 \times 10^{-22}$ & Fornax & $15$ &  $1.5$ & $N/A$ & 126.99 & $0.75$ & $0.45^{+0.07}_{-0.02}$ & $15.3^{+1.1}_{-3.4}$ & $80.2^{+17.2}_{-17.2}$ & $3.00$ & $136.42 $ & \\
$ 1 \times 10^{-22}$ & Fornax & $15$ &  $1.5$ & $N/A$ & 160.00 & $0.75$ & $0.52^{+0.00}_{-0.07}$ & $11.8^{+2.4}_{-1.4}$ & $98.6^{+18.9}_{-18.9}$ & $3.00$ & $136.42 $ & \\
$ 1 \times 10^{-22}$ & Fornax & $30$ &  $1.5$ & $N/A$ & 160.00 & $0.75$ & $0.44^{+0.06}_{-0.06}$ & $4.3^{+1.7}_{-0.5}$ & $149.2^{+18.9}_{-18.9}$ & $3.00$ & $0.00 $ & \\
$ 1 \times 10^{-22}$ & Fornax & $30$ &  $1.5$ & $N/A$ & 160.00 & $0.75$ & $0.64^{+0.01}_{-0.13}$ & $8.4^{+1.0}_{-1.0}$ & $112.6^{+30.1}_{-30.1}$ & $3.00$ & $50.02 $ & \ref{fig:1em22_Fornax} \\
$ 1 \times 10^{-22}$ & Fornax & $30$ &  $1.5$ & $N/A$ & 105.38 & $0.75$ & $0.46^{+0.02}_{-0.03}$ & $9.0^{+1.7}_{-0.5}$ & $101.7^{+9.4}_{-9.4}$ & $3.00$ & $0.00 $ & \\
$ 1 \times 10^{-22}$ & Fornax & $30$ &  $1.5$ & $N/A$ & 105.38 & $0.75$ & $0.59^{+0.06}_{-0.02}$ & $12.1^{+0.9}_{-2.3}$ & $85.9^{+10.0}_{-10.0}$ & $3.00$ & $50.02 $ & \\
$ 1 \times 10^{-22}$ & Fornax & $30$ &  $1.5$ & $N/A$ & 105.38 & $0.75$ & $0.57^{+0.09}_{-0.01}$ & $11.9^{+0.4}_{-1.7}$ & $85.2^{+13.0}_{-13.0}$ & $3.00$ & $50.02 $ & \\
$ 5 \times 10^{-22}$ & Fornax & $15$ &  $1.5$ & $N/A$ & 14.85 & $0.3$ & $1.12^{+0.05}_{-0.06}$ & $28.2^{+3.5}_{-5.0}$ & $9.1^{+1.3}_{-1.3}$ & $3.00$ & $45.47 $ & \\
$ 5 \times 10^{-22}$ & Fornax & $12$ &  $1.5$ & $N/A$ & 16.73 & $0.24$ & $0.80^{+0.05}_{-0.02}$ & $30.4^{+2.5}_{-7.0}$ & $9.6^{+1.0}_{-1.0}$ & $3.00$ & $68.21 $ & \\
$ 5 \times 10^{-22}$ & Fornax & $12$ &  $1.5$ & $N/A$ & 32.00 & $0.24$ & $0.32^{+0.02}_{-0.00}$ & $18.5^{+1.5}_{-2.3}$ & $16.6^{+0.9}_{-0.9}$ & $3.00$ & $68.21 $ & \\
$ 5 \times 10^{-22}$ & Fornax & $12$ &  $1.5$ & $1.71$ & N/A & $0.24$ & $0.77^{+0.04}_{-0.05}$ & $24.2^{+3.6}_{-1.8}$ & $11.5^{+1.1}_{-1.1}$ & $N/A$ & $50.02 $ & \ref{fig:5em22_Fornax_tides} \\
$ 5 \times 10^{-22}$ & Fornax & $12$ &  $1.5$ & $N/A$ & 32.00 & $0.24$ & $0.31^{+0.01}_{-0.00}$ & $17.6^{+2.0}_{-1.5}$ & $16.1^{+0.4}_{-0.4}$ & $3.00$ & $68.21 $ &\ref{fig:5em22_Fornax_tides} \\
$ 5 \times 10^{-22}$ & Fornax & $12$ &  $1.5$ & $N/A$ & 32.00 & $0.24$ & $0.33^{+0.01}_{-0.02}$ & $18.5^{+1.4}_{-1.7}$ & $15.6^{+0.7}_{-0.7}$ & $3.00$ & $68.21 $ & \\
$ 5 \times 10^{-22}$ & Fornax & $12$ &  $1.5$ & $1.05$ & N/A & $0.24$ & $0.63^{+0.04}_{-0.04}$ & $20.7^{+3.1}_{-1.4}$ & $14.8^{+1.1}_{-1.1}$ & $N/A$ & $50.02 $ & \\
$ 5 \times 10^{-22}$ & Fornax & $12$ &  $1.5$ & $1.05$ & N/A & $0.24$ & $0.91^{+0.01}_{-0.06}$ & $17.9^{+3.0}_{-2.3}$ & $15.4^{+1.9}_{-1.9}$ & $N/A$ & $0.00 $ & \ref{fig:5em22_Fornax_tides} \\
$ 1 \times 10^{-21}$ & Fornax & $12$ &  $1.5$ & $N/A$ & 12.70 & $0.24$ & $0.62^{+0.03}_{-0.03}$ & $25.7^{+0.3}_{-4.5}$ & $5.9^{+0.5}_{-0.5}$ & $3.00$ & $45.47 $ & \ref{fig:Fornax_1em21}, \ref{fig:snaps} \\
$ 2 \times 10^{-21}$ & Leo I/Draco & $4$ &  $1.0$ & $N/A$ & 6.35 & $0.04$ & $0.36^{+0.03}_{-0.01}$ & $18.7^{+3.3}_{-4.1}$ & $3.2^{+0.3}_{-0.3}$ & $1.50$ & $5.12 $ & \ref{fig:LeoI_2em21} \\
$ 2 \times 10^{-21}$ & Leo I/Draco & $4$ &  $1.0$ & $N/A$ & 6.35 & $0.04$ & $0.33^{+0.02}_{-0.01}$ & $21.2^{+0.7}_{-6.2}$ & $3.5^{+0.2}_{-0.2}$ & $1.00$ & $5.12 $ & \ref{fig:LeoI_2em21} \\
$ 5 \times 10^{-21}$ & Leo I/Draco & $4$ &  $1.0$ & $N/A$ & 3.79 & $0.04$ & $0.26^{+0.02}_{-0.02}$ & $14.3^{+1.9}_{-1.3}$ & $1.7^{+0.3}_{-0.3}$ & $1.50$ & $5.12 $ & \\
$ 5 \times 10^{-21}$ & Leo I/Draco* & $4$ &  $1.0$ & $N/A$ & 2.36 & $0.04$ & $0.31^{+0.00}_{-0.03}$ & $18.7^{+3.8}_{-2.2}$ & $1.1^{+0.2}_{-0.2}$ & $1.50$ & $5.12 $ & \\
$ 5 \times 10^{-21}$ & Leo I/Draco* & $4$ &  $1.0$ & $N/A$ & 2.02 & $0.04$ & $0.28^{+0.02}_{-0.01}$ & $23.7^{+2.9}_{-4.4}$ & $0.8^{+0.1}_{-0.1}$ & $1.50$ & $5.12 $ & \ref{fig:LeoI_5em21} \\
$ 5 \times 10^{-21}$ & Leo I/Draco* & $4$ &  $1.0$ & $N/A$ & 2.54 & $0.04$ & $0.25^{+0.01}_{-0.00}$ & $21.5^{+1.3}_{-0.1}$ & $1.0^{+0.0}_{-0.0}$ & $1.00$ & $5.12 $ & \ref{fig:LeoI_5em21} \\
$ 5 \times 10^{-21}$ & Leo II/Carina & $5$ &  $0.7$ & $1.83$ & N/A & $0.1$ & $0.27^{+0.02}_{-0.02}$ & $13.1^{+1.1}_{-0.9}$ & $1.9^{+0.2}_{-0.2}$ & $N/A$ & $5.46 $ & \\
$ 5 \times 10^{-21}$ & Leo II/Carina & $5$ &  $0.7$ & $1.18$ & N/A & $0.05$ & $0.33^{+0.02}_{-0.01}$ & $10.2^{+1.4}_{-0.4}$ & $2.7^{+0.4}_{-0.4}$ & $N/A$ & $1.09 $ & \ref{fig:leoII} \\
$ 5 \times 10^{-21}$ & Leo II/Carina & $5$ &  $0.7$ & $1.83$ & N/A & $0.05$ & $0.21^{+0.02}_{-0.01}$ & $12.5^{+0.8}_{-1.2}$ & $2.3^{+0.3}_{-0.3}$ & $N/A$ & $5.46 $ & \ref{fig:leoII} \\
$ 5 \times 10^{-21}$ & Leo II/Carina & $5$ &  $0.7$ & $1.75$ & N/A & $0.05$ & $0.29^{+0.02}_{-0.01}$ & $9.6^{+1.6}_{-0.3}$ & $2.6^{+0.3}_{-0.3}$ & $N/A$ & $2.18 $ & \ref{fig:leoII} \\
$ 5 \times 10^{-21}$ & Leo II/Carina & $5$ &  $0.7$ & $N/A$ & 4.78 & $0.05$ & $0.29^{+0.02}_{-0.01}$ & $10.0^{+0.3}_{-1.9}$ & $3.0^{+0.3}_{-0.3}$ & $3.00$ & $1.09 $ & \\
$ 5 \times 10^{-21}$ & Leo II/Carina & $5$ &  $0.7$ & $0.87$ & N/A & $0.05$ & $0.34^{+0.01}_{-0.02}$ & $7.5^{+0.4}_{-0.5}$ & $3.7^{+0.3}_{-0.3}$ & $N/A$ & $1.09 $ & \\
$ 5 \times 10^{-21}$ & Leo II/Carina & $5$ &  $0.7$ & $1.75$ & N/A & $0.05$ & $0.28^{+0.01}_{-0.02}$ & $11.0^{+0.5}_{-0.6}$ & $2.4^{+0.3}_{-0.3}$ & $N/A$ & $2.18 $ & \\
$ 5 \times 10^{-21}$ & Leo II/Carina & $5$ &  $0.7$ & $N/A$ & 4.78 & $0.05$ & $0.38^{+0.02}_{-0.02}$ & $8.6^{+1.4}_{-0.8}$ & $2.9^{+0.5}_{-0.5}$ & $2.00$ & $1.09 $ & \\
$ 5 \times 10^{-21}$ & Leo II/Carina & $5$ &  $0.7$ & $N/A$ & 4.78 & $0.05$ & $0.32^{+0.03}_{-0.01}$ & $8.5^{+2.3}_{-0.6}$ & $2.6^{+0.4}_{-0.4}$ & $2.00$ & $1.09 $ & \ref{fig:LeoII_5em21_tides} \\
$ 5 \times 10^{-21}$ & Leo II/Carina & $5$ &  $0.7$ & $N/A$ & 4.78 & $0.05$ & $0.25^{+0.01}_{-0.02}$ & $10.8^{+1.1}_{-1.3}$ & $2.5^{+0.2}_{-0.2}$ & $1.00$ & $1.09 $ & \ref{fig:LeoII_5em21_tides} \\
$ 5 \times 10^{-21}$ & Leo II/Carina & $5$ &  $0.7$ & $N/A$ & 4.78 & $0.05$ & $0.14^{+0.00}_{-0.01}$ & $10.2^{+0.8}_{-2.4}$ & $2.8^{+0.1}_{-0.1}$ & $0.50$ & $1.09 $ & \ref{fig:LeoII_5em21_tides} \\
$ 5 \times 10^{-21}$ & Leo II/Carina & $5$ &  $0.7$ & $1.75$ & N/A & $0.05$ & $0.41^{+0.01}_{-0.01}$ & $7.4^{+1.7}_{-0.2}$ & $3.2^{+0.6}_{-0.6}$ & $N/A$ & $0.00 $ & \ref{fig:leoII} \\
$ 5 \times 10^{-21}$ & Leo II/Carina & $5$ &  $0.7$ & $1.75$ & N/A & $0.05$ & $0.34^{+0.02}_{-0.01}$ & $9.7^{+1.1}_{-2.4}$ & $2.8^{+0.4}_{-0.4}$ & $N/A$ & $1.09 $ & \\
$ 5 \times 10^{-21}$ & Leo I/Draco* & $4$ & $0.8$& $N/A$ & 2.50 & $0.04$ & $0.19^{+0.01}_{-0.01}$ & $14.9^{+4.7}_{-1.5}$ & $0.9^{+0.1}_{-0.1}$ & $1.00$ & $5.00 $ &  \\
$ 5 \times 10^{-21}$ & Leo I/Draco* & $4$ & $1.2$& $N/A$ & 2.50 & $0.04$ & $0.19^{+0.01}_{-0.01}$ & $22.4^{+5.6}_{-4.6}$ & $0.9^{+0.0}_{-0.0}$ & $1.00$ & $5.00 $ &  \\
$ 5 \times 10^{-21}$ & Leo I/Draco* & $4$ & $0.8$& $N/A$ & 3.50 & $0.04$ & $0.11^{+0.01}_{-0.01}$ & $20.8^{+2.6}_{-1.7}$ & $1.3^{+0.1}_{-0.1}$ & $1.00$ & $5.00 $ &  \\
$ 5 \times 10^{-21}$ & Leo I/Draco*$^\dagger$ & $4$ & $1.0$& $N/A$ & 3.00 & $0.04$ & $0.22^{+0.01}_{-0.02}$ & $24.3^{+3.0}_{-6.5}$ & $0.9^{+0.1}_{-0.1}$ & $1.00$ & $5.00 $ &  \\
$ 5 \times 10^{-21}$ & Leo I/Draco*$^\top$ & $4$ & $1.0$& $N/A$ & 2.50 & $0.04$ & $0.20^{+0.01}_{-0.01}$ & $24.0^{+8.9}_{-2.4}$ & $0.7^{+0.0}_{-0.0}$ & $1.00$ & $5.00 $ &  \\
$ 2 \times 10^{-21}$ & Leo I/Draco* & $10$ & $1.0$& $N/A$ & 6.00 & $0.04$ & $0.30^{+0.02}_{-0.01}$ & $22.5^{+2.1}_{-1.4}$ & $2.8^{+0.1}_{-0.1}$ & $1.00$ & $5.00 $ &  \\
$ 5 \times 10^{-21}$ & Leo II/Carina* & $10$ & $0.7$& $N/A$ & 5.00 & $0.05$ & $0.41^{+0.02}_{-0.02}$ & $9.9^{+2.8}_{-0.4}$ & $2.1^{+0.2}_{-0.2}$ & $2.00$ & $1.00 $ &  \\
\bottomrule
\end{tabular}
\caption{Overview of our simulations suite. 
{\bf Column 1}: particle mass in eV. 
{\bf Column 2}: target system, asterisks denote high-resolution simulations with number of grid points $N=512^3$ instead of $N=256^3$.
{\bf Column 3}: Length of the box in kpc. 
{\bf Column 4}: $r_{\rm s} $ in kpc. 
{\bf Column 5}: NFW $\rho_{\rm s}$ parameter for simulations initialized with Eddington, in $M_\odot/$kpc$^3$, applicable only for simulations without tides (if with tides, we use a different method, where the normalization of the profile is specified by the parameter reported on column 6). 
{\bf Column 6}: $r_{\rm c}$ used to initialize the initial soliton profile, in $10$ pc; applicable only for simulations where tides are explicitly taken into account (as explained in Sec.~\ref{s:sims}).
{\bf Column 7}: initial $r_{\rm plum} $ for the stars distribution in kpc. 
{\bf Column 8}: $r_{\rm half}$ in kpc after $10$ Gyr, error bars show the spread for $10\pm 1$ Gyr. 
{\bf Column 9}: $\sigma_{\rm los}$ after $10$ Gyr. in the innermost bin in km/s, error bars show the spread for $10\pm 1$ Gyr.
{\bf Column 10}: core radius of the soliton in $10$ pc, error bars show the spread for $10\pm 1$ Gyr.
{\bf Column 11}: tidal radius in kpc (if used).
{\bf Column 12}: total mass of the stars in $10^6 M_\odot$.
{\bf Column 13}: figures in which the simulation results appear.
$\dagger$: transition radius parameter $r_{\rm r} = 2$; $\top$: transition radius parameter $r_{\rm r} = 3$.}
\label{tab:details}
\end{table*}

\bibliographystyle{bibi}
\bibliography{biblio}

\end{document}